\documentclass[nofootinbib,aps,amssymb,floatfix,superscriptaddress,preprintnumbers]{revtex4}

\usepackage{graphicx}
\usepackage{bm}
\usepackage{amsmath}
\usepackage{amssymb}
\usepackage{amsfonts}
\usepackage{float}
\usepackage{hyperref}
\usepackage{dsfont}  
\usepackage{slashed}  
\usepackage{booktabs}
\usepackage{multirow}
\usepackage{subfigure}
\usepackage[sort&compress]{natbib}
\usepackage{xcolor}
\usepackage{ulem}

\newcommand{\be}{\begin{equation}}  
\newcommand{\ee}{\end{equation}}  
\newcommand{\beq}{\begin{eqnarray}} 
\newcommand{\eeq}{\end{eqnarray}}

\newcommand{\bea}{\begin{eqnarray}}
\newcommand{\eea}{\end{eqnarray}}

\newcommand{\Slash}[1]{{\ooalign{\hfil/\hfil\crcr$#1$}}}
\newcommand{\nn}{\nonumber \\}

\begin{document}
\preprint{LA-UR-24-22343}

\title{Exploring orbital angular momentum and spin-orbit correlation for gluons \\ at the Electron-Ion Collider }

\author{Shohini Bhattacharya}
\email{shohinib@lanl.gov}
\affiliation{Theoretical Division, Los Alamos National Laboratory, Los Alamos, New Mexico 87545, USA}
\affiliation{RIKEN BNL Research Center, Brookhaven National Laboratory, Upton, NY 11973, USA}

\author{Renaud Boussarie}
\email{renaud.boussarie@polytechnique.edu}
\affiliation{CPHT, CNRS, Ecole Polytechnique, Institut Polytechnique de Paris, 91128 Palaiseau, France}

\author{Yoshitaka Hatta}
\email{yhatta@bnl.gov}
\affiliation{Physics Department, Brookhaven National Laboratory, Upton, NY 11973, USA}
\affiliation{RIKEN BNL Research Center, Brookhaven National Laboratory, Upton, NY 11973, USA}

\newcommand{\Shohini}[1]{{\color{blue}\textbf{SB}: {#1}}}

\begin{abstract}
In our previous work~\cite{Bhattacharya:2022vvo}, we introduced a pioneering observable aimed at experimentally detecting  the orbital angular momentum (OAM) of gluons.  Our focus was on the longitudinal double spin asymmetry observed in exclusive dijet production during electron-proton scattering. We demonstrated the sensitivity of the $\cos \phi$ angular correlation between the scattered electron and proton as a probe for gluon OAM at small-$x$ and its intricate interplay with gluon helicity. This current work provides a comprehensive exposition, diving further into the aforementioned calculation with added elaboration and in-depth analysis. We reveal that, in addition to the gluon OAM, one also gains access to the spin-orbit correlation of gluons.  
We supplement our work with a detailed numerical analysis of our observables for the kinematics of the Electron-Ion Collider. In addition to dijet production, we  also consider the recently proposed semi-inclusive diffractive deep inelastic scattering process which potentially offers  experimental advantages over dijet measurements. 
Finally, we investigate quark-channel contributions to these processes and find an unexpected breakdown of collinear factorization. 

\end{abstract}

\maketitle

\section{Introduction}
\label{sec:introduction}
The investigation of nucleon spin structure, prompted by the discovery of the ``spin crisis", has evolved into a captivating research field over the past three decades. A primary objective of this field is to comprehend the spin decomposition of the nucleon. As per the Jaffe-Manohar decomposition \cite{Jaffe:1989jz}, the proton spin originates from four distinct sources
\begin{eqnarray}
\frac{1}{2}=\frac{1}{2}\Delta \Sigma+\Delta G+L_q+L_g \, .
\end{eqnarray}
The contribution of quark spin, $\Delta \Sigma \sim 0.3$, obtained from the integration of the quark helicity distribution, is relatively well constrained. However, the gluon spin contribution, $\Delta G$, is anticipated to be significant~\cite{STAR:2014wox,deFlorian:2014yva,Nocera:2014gqa,STAR:2021mqa}, although the uncertainties surrounding the gluon helicity distribution at small $x$ are substantial. The forthcoming Electron-Ion Collider (EIC) is expected to provide precise experimental data that will allow for a more accurate determination of the gluon spin contribution~\cite{AbdulKhalek:2021gbh}.

In addition to the ongoing exploration of the gluon helicity distribution at small $x$, there is currently significant attention devoted to investigating the orbital angular momentum (OAM) contributions, $L_{q/g}$, from quarks and gluons. This pursuit holds great promise for advancing our understanding of the partonic structure of the nucleon and shedding light on the underlying dynamics of QCD. To date, our knowledge regarding the OAM of partons remains limited despite much progress in formal theory~\cite{Lorce:2011kd,Hatta:2011ku,Lorce:2011ni,Hatta:2012cs,Ji:2012ba,Rajan:2016tlg,Engelhardt:2017miy,Boussarie:2019icw,Guo:2021aik,Kovchegov:2019rrz,Engelhardt:2020qtg,Kovchegov:2023yzd,Manley:2024pcl}. The extraction of Jaffe-Manohar-type (or canonical) parton OAM from high-energy scattering experiments remains highly challenging, both from theoretical and experimental perspectives. Previous approaches for probing parton OAM have relied on its connection with the Wigner distribution, or equivalently, the generalized transverse momentum distribution function (GTMD)~\cite{Lorce:2011kd,Hatta:2011ku,Lorce:2011ni}, 
\begin{eqnarray} 
xL_{g}(x,\xi)=-\int d^2 k_\perp \frac{ 
 k_\perp^2}{M^2} F_{1,4}^{g}(x, k_\perp,\xi, \Delta_\perp=0) \, ,
\end{eqnarray}
where the specific kinematical variables will be provided below. (For quarks, the above equation remains the same, except that there is no factor of $x$ on the left-hand side.) The parton OAM can be obtained by integrating the $x$-dependent OAM distribution: $L_{g,q}=\int_0^1 dx L_{g,q}(x,\xi=0)$. Significant theoretical advancements have been made in investigating the experimental signatures of the GTMD $F_{1,4}$ in recent years. In Refs.~\cite{Ji:2016jgn,Hatta:2016aoc,Bhattacharya:2022vvo}, it was demonstrated that the single and double longitudinal spin asymmetry observed in exclusive dijet production during $ep$ collisions provides access to the ``Compton form factor" (or $x$-moments) involving $F_{1,4}$. Furthermore, the direct measurement of $F_{1,4}$, along with several other GTMDs, can be accomplished through the exclusive double Drell-Yan or quarkonium production~\cite{Bhattacharya:2017bvs,Bhattacharya:2018lgm,Boussarie:2018zwg}. An earlier notable endeavor aimed at identifying an observable for probing OAM can be found in Ref.~\cite{Courtoy:2013oaa,Rajan:2016tlg}. A recent study proposed a method to extract the imaginary component of the gluon GTMD $F_{1,4}$ by measuring the azimuthal angular correlation of $\sin 2\phi$ in exclusive $\pi^0$ production during polarized $ep$ collisions~\cite{Bhattacharya:2023yvo}. This approach exploits the interference effect between the QCD interaction and the Primakoff process, resulting in a single-target longitudinal spin asymmetry. Additionally, in Ref.~\cite{Bhattacharya:2023hbq}, the same $\sin 2\phi$ asymmetry, which mirrors the aforementioned but does not arise from the interference effect described there, provides access to the quark GTMD $F_{1,4}$. Hence, there has been a consistent advancement in identifying observables that exhibit sensitivity to OAM.

In our earlier work~\cite{Bhattacharya:2022vvo}, we introduced a novel observable to experimentally detect the OAM of gluons, which, as discussed above, plays a pivotal role in the proton spin sum rule. We focused on the longitudinal double spin asymmetry observed in exclusive dijet production during $ep$ scattering, demonstrating that the $\cos \phi$ angular correlation between the scattered electron and proton serves as a probe of gluon OAM at small-$x$, as well as its intricate interplay with gluon helicity. In this study, we provide a comprehensive exposition, offering further elaboration and in-depth analysis of the aforementioned calculation. We reveal that, in addition to gluon OAM and its helicity, one also gains access to the spin-orbit correlation of gluons. 
We then provide a detailed numerical analysis of our observable in the dijet process. Due to practical complications associated with dijet reconstruction, we also investigate the possibility to reinterpret our observable in terms of  semi-inclusive diffractive deep inelastic scattering (`SIDDIS') \cite{Hatta:2022lzj} and perform a numerical analysis. Furthermore, we undertake the calculation of quark-channel contributions to these processes. Contrary to our initial expectations, we discover an unexpected breakdown of factorization compared to the case of the gluon channel.

The paper is structured as follows: In Sec.~\ref{s:def}, we establish the definitions of GTMDs for both gluons and quarks, elucidating their connections with OAM. Additionally, we briefly explain the subtlety associated with the  frame-dependence of GTMDs. The primary findings regarding exclusive dijet production are detailed in Sec.~\ref{s:main}, where we discuss the calculation of the observable double spin asymmetry. 
In Sec.~\ref{s:quark_channel}, we explore the quark channel contribution to the dijet process, addressing factors contributing to the breakdown of factorization. Sec.~\ref{s:siddis} explores the semi-inclusive diffractive deep inelastic scattering (SIDDIS) process, similar to dijets but with a focus on tagging inclusive hadron species within a specific rapidity interval. The final numerical results for both dijets and SIDDIS under realistic EIC kinematics are shown in Sec.~\ref{s:numerics}. Conclusions are presented in Sec.~\ref{s:conclusion}.

\section{GTMDs and orbital angular momentum }
\label{s:def}
\subsection{Definitions}
In this section we recapitulate  the basic definition of GTMDs and their connection to the parton orbital angular momentum (OAM). 
Following \cite{Meissner:2009ww}, we parameterize the leading-twist quark and gluon GTMDs as\footnote{Our conventions are $\gamma_5=-i\gamma^0\gamma^1\gamma^2\gamma^3$ and $\epsilon^{0ij3}=\epsilon^{-+ij}=\epsilon^{ij}=\epsilon_{ij}$ ($i,j=1,2$) such that ${\rm Tr}[\gamma_5\gamma^\mu \gamma^\nu \gamma^\rho\gamma^\lambda]=4i\epsilon^{\mu\nu\rho\lambda}$ and  the spin four-vector is $2s^\mu = \bar{u}(ps)\gamma_5\gamma^\mu u(ps)$. }

\beq
f_q(x,\xi,\widetilde{k}_\perp,\widetilde{\Delta}_\perp)&=&\int \frac{d^3z}{2(2\pi)^3} e^{ixP^+z^--i\widetilde{k}_\perp\cdot \widetilde{z}_\perp}\langle p'|\bar{\psi}(-z/2)\gamma^+\psi(z/2)|p\rangle  \nn 
&=& \frac{1}{2M}\bar{u}(p')\left[ F_{1,1}^q+i\frac{\sigma^{j+}}{P^+}(\widetilde{k}_\perp^j F_{1,2}^q+\widetilde{\Delta}_\perp^j F^q_{1,3})+ i\frac{\sigma^{ij}\widetilde{k}_\perp^i\widetilde{\Delta}_\perp^j}{M^2}F^q_{1,4}\right]u(p), \label{gtmd1} 
\eeq
\beq
\tilde{f}_q(x,\xi,\widetilde{k}_\perp,\widetilde{\Delta}_\perp)&=& \int \frac{d^3z}{2(2\pi)^3} e^{ixP^+z^--i\widetilde{k}_\perp\cdot \widetilde{z}_\perp}\langle p'|\bar{\psi}(-z/2)\gamma_5\gamma^+(z/2)|p\rangle \nn 
&=& \frac{-i}{2M}\bar{u}(p')\left[ \frac{\epsilon_{ij}\widetilde{k}_\perp^i \widetilde{\Delta}_\perp^j}{M^2}G^q_{1,1}-\frac{\sigma^{i+}\gamma_5}{P^+}(\widetilde{k}_\perp^i G^q_{1,2}+\widetilde{\Delta}_\perp^i G^q_{1,3})-\sigma^{+-}\gamma_5G^q_{1,4}\right]u(p), \label{gtmdqtilde} \\[0.4cm]
xf_g(x,\xi,\widetilde{k}_\perp,\widetilde{\Delta}_\perp)&=&\delta_{ij}\int \frac{d^3z}{(2\pi)^3P^+} e^{ixP^+z^--i\widetilde{k}_\perp\cdot \widetilde{z}_\perp}\langle p'|F_a^{+i}(-z/2)F_a^{+j}(z/2)|p\rangle  \nn 
&=& \frac{1}{2M}\bar{u}(p')\left[ F^g_{1,1}+i\frac{\sigma^{j+}}{P^+}(\widetilde{k}_\perp^j F^g_{1,2}+\widetilde{\Delta}_\perp^j F^g_{1,3})+ i\frac{\sigma^{ij}\widetilde{k}_\perp^i\widetilde{\Delta}_\perp^j}{M^2}F^g_{1,4}\right]u(p), \label{f14}
\\[0.4cm]
x\tilde{f}_g(x,\xi,\widetilde{k}_\perp,\widetilde{\Delta}_\perp)&=& -i\epsilon_{ij}\int \frac{d^3z}{(2\pi)^3P^+} e^{ixP^+z^--i\widetilde{k}_\perp\cdot \widetilde{z}_\perp}\langle p'|F_a^{+i}(-z/2)F_a^{+j}(z/2)|p\rangle \nn 
&=& \frac{-i}{2M}\bar{u}(p')\left[ \frac{\epsilon_{ij}\widetilde{k}_\perp^i \widetilde{\Delta}_\perp^j}{M^2}G^g_{1,1}-\frac{\sigma^{i+}\gamma_5}{P^+}(\widetilde{k}_\perp^i G^g_{1,2}+\widetilde{\Delta}_\perp^i G^g_{1,3})-\sigma^{+-}\gamma_5 G^g_{1,4}\right]u(p), \label{gtmd2}
\eeq 
where $P^\mu = \frac{p+p'}{2}$, $\Delta^\mu=p'^\mu- p^\mu$ and $\xi=(p^+-p'^+)/2P^+$. $i,j=1,2$ are two-dimensional vector indices. For the quark GTMDs, a staple-shaped gauge link $W^\pm$ which connects the two points $-z/2$ and $z/2$ via the light-cone infinity $z^-=\pm\infty$  is understood. For the gluon GTMDs, there are two inequivalent ways to insert Wilson lines 
\beq
F_a(-z/2)F_a(z/2) \to \begin{cases}  2{\rm tr}[F(-z/2)W^\pm F(z/2) (W^\pm)^\dagger], \\  2{\rm tr}[F(-z/2)W^\pm F(z/2) (W^\mp)^\dagger], \end{cases} 
\label{wwdipole}
\eeq
where the matrices are in the fundamental representation. The first choice is called the Weisz\"acker-Williams gluon distribution and the second is the dipole gluon distribution. Which gluon distribution one should adopt depends on the process under consideration \cite{Bomhof:2006dp,Dominguez:2011wm}.  For exclusive dijet production, we select the dipole one \cite{Altinoluk:2015dpi,Hatta:2016dxp,Boussarie:2016ogo}. 

All the GTMDs are a function of $x,\xi,\widetilde{k}_\perp^2,\widetilde{\Delta}_\perp^2$ and $\widetilde{k}_\perp\cdot \widetilde{\Delta}_\perp$. (Throughout this paper we denote $A_\perp^2=A_\perp^i A_\perp^i$, $A_\perp\cdot B_\perp = A_\perp^i B_\perp^i$, and $A_\perp \times B_\perp = \epsilon^{ij}A_\perp^i B_\perp^j$ for products of two-dimensional vectors.) 
The twist-two quark and gluon GPDs are obtained by integrating over $\widetilde{k}_\perp$. 
\beq
\int d^2\widetilde{k}_\perp f_q &=& \frac{1}{2P^+} \bar{u}(p')\left(H_q\gamma^++ E_q\frac{i\sigma^{+\nu}\Delta_\nu}{2M}\right)u(p),  
\eeq
\beq
\int d^2\widetilde{k}_\perp \tilde{f}_q &=& \frac{1}{2P^+} \bar{u}(p')\left(\tilde{H}_q\gamma_5\gamma^+- \tilde{E}_q\frac{\gamma_5\Delta^+}{2M}\right)u(p), \label{quarkgluonoam}
\eeq
\beq
\int d^2\widetilde{k}_\perp xf_g &=& \frac{1}{2P^+} \bar{u}(p')\left(H_g\gamma^++ E_g\frac{i\sigma^{+\nu}\Delta_\nu}{2M}\right)u(p), \label{hg}
\eeq
\beq
\int d^2\widetilde{k}_\perp x\tilde{f}_g &=& \frac{1}{2P^+} \bar{u}(p')\left(\tilde{H}_g\gamma_5\gamma^+- \tilde{E}_g\frac{\gamma_5\Delta^+}{2M}\right)u(p).
\label{hgtilde}
\eeq
In the forward limit, they reduce to the unpolarized and polarized quark and gluon PDFs as $H_q(x)=q(x)$, $\tilde{H}_q(x)=\Delta q(x)$, $H_g(x)=xG(x)$ and $\tilde{H}_g(x)=x\Delta G(x)$. 
Let us define
\beq
L_q (x,\xi )\equiv  -\int d^2\widetilde{k}_\perp \frac{\widetilde{k}_\perp^2}{M^2}F^q_{1,4}(x,\xi,\widetilde{\Delta}_\perp=0), \qquad xL_g (x,\xi )\equiv  -\int d^2\widetilde{k}_\perp \frac{\widetilde{k}_\perp^2}{M^2}F^g_{1,4}(x,\xi,\widetilde{\Delta}_\perp=0).
\label{lgx0}
\eeq
 In the limit $\xi\to 0$, these become the quark and gluon OAM parton distribution functions \cite{Hatta:2012cs}
 \beq
L_q (x)= -\int d^2\widetilde{k}_\perp \frac{\widetilde{k}_\perp^2}{M^2}F^q_{1,4}(x,\xi=0,\widetilde{\Delta}_\perp=0), \qquad xL_g (x)= -\int d^2\widetilde{k}_\perp \frac{\widetilde{k}_\perp^2}{M^2}F^g_{1,4}(x,\xi=0,\widetilde{\Delta}_\perp=0),
\label{lgx}
\eeq
 normalized as\footnote{Our normalization of $L_g(x)$ differs from the one in  \cite{Hatta:2012cs} by a factor of 2.} 
\beq
\int_{-1}^1dx L_q(x) = \int_0^1dx (L_q(x)+L_{\bar{q}}(x)) = L_q, \qquad \int_0^1dx L_g(x)=L_g. 
\eeq
Due to $PT$-symmetry, $L_{g,q}(x)$ does not depend on the direction (future-pointing or past-pointing) of the gauge link $W^\pm$ \cite{Hatta:2011ku}. Moreover, in the gluon case the  two choices (\ref{wwdipole}) result in the same function $L_{g}(x)$ \cite{Hatta:2016aoc}. 
$L_{q,g}(x)$ can be decomposed into the Wandzura-Wilczek part related to the twist-two PDFs and the genuine twist-three part (matrix elements of $qqg$ and $ggg$ correlators)
\beq
L_q(x) &=& x\int_x^{\epsilon(x)}\frac{dx'}{x'}(H_q(x')+E_q(x'))-x\int_x^{\epsilon(x)} \frac{dx'}{x'^2}\tilde{H}_q(x')+\cdots, \qquad (-1<x<1)\\ 
L_g(x) &=& x\int_x^1\frac{dx'}{x'^2} (H_g(x') +E_g(x'))-2x\int_x^1\frac{dx'}{x'^2}\Delta G(x')+\cdots, \qquad (0<x<1) \label{www}
\eeq
where $\epsilon(x)$ is the sign function. The genuine  twist-three parts are omitted for simplicity. Their complete expressions can be found in  \cite{Hatta:2012cs,Hatta:2019csj}.  
At small-$x$, the parton distributions have the power-law behavior 
\beq
H_g(x)\sim E_g(x)\sim \frac{1}{x^c}, \qquad \Delta G(x)\sim \frac{1}{x^d}.\label{bfkl}
\eeq
[The result $H_g(x)\propto E_g(x)$ at small-$x$ has been recently established in \cite{Hatta:2022bxn}.] 
In perturbation theory, $c\propto \alpha_s$ \cite{Kuraev:1977fs,Balitsky:1978ic} and $d\propto \sqrt{\alpha_s}$ \cite{Bartels:1996wc,Borden:2023ugd}. We thus expect that $c<d$ and   the right hand side of (\ref{www}) is dominated by the $\Delta G$  term \cite{Boussarie:2019icw,Kovchegov:2023yzd,Manley:2024pcl}
\beq
L_g(x) \approx -\frac{2}{1+d}\Delta G(x).\label{lww} 
\eeq
This relation will play an important role in our analysis below. It means that a significant cancellation occurs between the OAM and helicity distributions at small-$x$ \cite{Hatta:2016aoc,More:2017zqp,Hatta:2018itc,Boussarie:2019icw}. Recently, it has been shown that this feature is robust even after including the genuine twist-three corrections \cite{Kovchegov:2023yzd,Manley:2024pcl} (see also \cite{Hatta:2019csj,Kovchegov:2019rrz}).

\subsection{Transverse boost}
It is often taken for granted that GTMDs are  defined in the `symmetric' frame where $P_\perp=0$ so that $\widetilde{p}_\perp' = \widetilde{\Delta}_\perp/2 = -\widetilde{p}_\perp$. The advantage of this frame is that one can exploit hermiticity and $PT$  (parity \& time-reversal) symmetry to constrain the dependence of GTMDs on various variables \cite{Meissner:2009ww}
\beq
&&\mathfrak{Re}\, F_{1,n}(x,\xi, \widetilde{k}_\perp\cdot \widetilde{\Delta}_\perp,\widetilde{k}_\perp^2,\widetilde{\Delta}_\perp^2) = \mathfrak{Re}\, F_{1,n}(x,-\xi,-\widetilde{k}_\perp \cdot \widetilde{\Delta}_\perp,\widetilde{k}_\perp^2,\widetilde{\Delta}_\perp^2),\\ && \mathfrak{Im}\, F_{1,n}(x,\xi, \widetilde{k}_\perp\cdot \widetilde{\Delta}_\perp,\widetilde{k}_\perp^2,\widetilde{\Delta}_\perp^2) = - \mathfrak{Im}\, F_{1,n}(x,-\xi,-\widetilde{k}_\perp \cdot \widetilde{\Delta}_\perp,\widetilde{k}_\perp^2,\widetilde{\Delta}_\perp^2) ,
\eeq
for $n=1,3,4$ and 
\beq
&&\mathfrak{Re}\, F_{1,2}(x,\xi, \widetilde{k}_\perp\cdot \widetilde{\Delta}_\perp,\widetilde{k}_\perp^2,\widetilde{\Delta}_\perp^2) = -\mathfrak{Re}\, F_{1,2}(x,-\xi,-\widetilde{k}_\perp \cdot \widetilde{\Delta}_\perp,\widetilde{k}_\perp^2,\widetilde{\Delta}_\perp^2), \label{real12}\\ && \mathfrak{Im}\, F_{1,2}(x,\xi, \widetilde{k}_\perp\cdot \widetilde{\Delta}_\perp,\widetilde{k}_\perp^2,\widetilde{\Delta}_\perp^2) =  \mathfrak{Im}\, F_{1,2}(x,-\xi,-\widetilde{k}_\perp \cdot \widetilde{\Delta}_\perp,\widetilde{k}_\perp^2,\widetilde{\Delta}_\perp^2).
\eeq
Throughout this paper, we are interested in the regime where $\xi$ and $\Delta_\perp$ are small. Then the above relations already imply that $F_{1,n=1,3,4}$ are dominated by the real part and $F_{1,2}$ is dominated by the imaginary part. In the forward limit, ${\mathfrak Im}F^{q/g}_{1,2}$ are called the quark/gluon Sivers function. 
However, this frame is neither convenient nor practically  used when describing actual experimental  processes. We shall instead work in the so-called hadron frame, where the incoming virtual photon and the proton move along the $-x^3$ and $+x^3$ directions, respectively, with $p_\perp=0$.
The two frames are related by the so-called {\it transverse boost}, a Lorentz transformation that leaves invariant the plus component of a four-vector $V^\mu=(V^+,V^-,V_\perp)$
\beq
V^+ = \widetilde{V}^+, \qquad 
V_\perp = \widetilde{V}_\perp + C_\perp \widetilde{V}^+, \qquad 
V^- = \widetilde{V}^- + C_\perp \cdot \widetilde{V}_\perp +\frac{C_\perp^2}{2}\widetilde{V}^+
\eeq
where $C_\perp$ is an arbitrary two-dimensional vector. 
We start with the definition of GTMDs in the symmetric frame (\ref{gtmd1})-(\ref{gtmd2}) and apply the following transverse boost with $C_\perp = \frac{\widetilde{\Delta}_\perp}{2p^+}$  
\beq
p^+,p'^+, &\to& p^+,p'^+,\nn
z^+=0 &\to& 0, \nn 
\widetilde{p}_\perp =-\frac{\widetilde{\Delta}_\perp}{2} &\to& p_\perp=0, \nn 
\widetilde{p}'_\perp = \frac{\widetilde{\Delta}_\perp}{2} &\to& p'_\perp =\frac{\widetilde{\Delta}_\perp}{2} + \frac{\widetilde{\Delta}_\perp}{2p^+}p'^+ = \frac{\widetilde{\Delta}_\perp}{1+\xi} \equiv \Delta_\perp, \nn
 z_\perp &\to&  z_\perp,\nn
z^- &\to&   z^- + \frac{z_\perp \cdot \widetilde{\Delta}_\perp}{2p^+} + {\cal O}(\Delta^2_\perp) .
\eeq
Then in (\ref{gtmd1})-(\ref{gtmd2}) one can write
\beq
\int dz^-d^2z_\perp e^{ixP^+\left(z^- -\frac{z_\perp\cdot \widetilde{\Delta}_\perp}{2p^+}\right) -i\widetilde{k}_\perp \cdot z_\perp}\bigl\langle (1-\xi)P^+,\Delta_\perp\bigr|\cdots \left|(1+\xi)P^+,0_\perp \right\rangle , \nn 
= \int dz^-d^2z_\perp e^{ixP^+z^- -i\left(\widetilde{k}_\perp +\frac{x \widetilde{\Delta}_\perp}{2(1+\xi)}\right)\cdot z_\perp  }\bigl\langle (1-\xi)P^+,\Delta_\perp \bigr|\cdots \left|(1+\xi)P^+,0_\perp \right\rangle .
\eeq
This means that if one considers scattering processes in the hadron frame where $p_\perp=0$, transverse momentum transfer $p'_\perp=\Delta_\perp$ and $t$-channel gluons with transverse momentum $k_\perp$, the GTMDs $F_{1,n}(x,\xi,\widetilde{k}_\perp,\widetilde{\Delta}_\perp)$ should be evaluated at 
\beq
&& \widetilde{k}_\perp = k_\perp  -\frac{x\widetilde{\Delta}_\perp}{2(1+\xi)} = k_\perp -\frac{x}{2}\Delta_\perp, \qquad  \widetilde{\Delta}_\perp= (1+\xi)\Delta_\perp. \label{difference}
\eeq
Many of the previous phenomenological applications of GTMDs, including this work, have focused on the small-$x$ kinematics $x\ll 1$, or more precisely, $\xi\ll 1$. Moreover, in the subsequent calculations we shall only keep the linear terms in $\Delta_\perp$, and the ${\cal O}(\Delta_\perp)$ shift in $\widetilde{k}_\perp$ turns out to be a higher order effect. Therefore, for the purpose of the present work, the differences  (\ref{difference}) can be neglected. We however envisage more general reactions involving GTMDs where the difference must be taken into account.  
We finally note that the longitudinally polairized spin vector of the incoming proton 
\beq
s^\mu=(s^+,s^-,0_\perp) = h_p\left( (1+\xi)P^+, \frac{-M^2}{2(1+\xi)P^+},0_\perp\right),
\eeq
in the hadron frame corresponds in the symmetric frame to 
\beq
\widetilde{s}^\mu = \left(s^+,s^- +\frac{\Delta_\perp^2 s^+}{8(p^+)^2}, -\frac{s^+}{2p^+}\Delta_\perp  \right) .
\eeq

\section{Double spin asymmetry in exclusive dijet production}
\label{s:main}
\subsection{General considerations} 
\begin{figure}
  \includegraphics[width=0.5\linewidth]{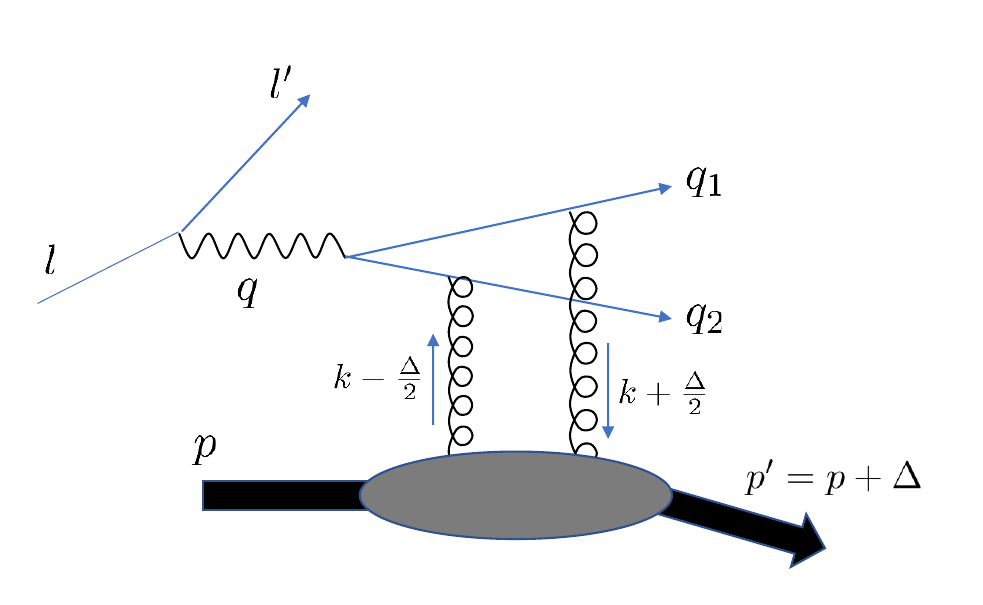}
\caption{Exclusive dijet production in electron-proton scattering.}
\label{main_dia}
\end{figure} 

In this section, we calculate the cross section of longitudinal double spin asymmetry in exclusive dijet production. A brief summary of the result has been reported in our  previous publication \cite{Bhattacharya:2022vvo}. The process is depicted in Fig.~\ref{main_dia}. At high center-of-mass energies, the dominant contribution comes from the two-gluon exchange diagrams. The quark exchange diagrams will be discussed in a later section.  

We assume that both the incoming proton and lepton are longitudinally polarized with helicity $h_p=\pm 1$ and $h_l=\pm 1$, respectively. We write the cross section $d\sigma^{h_p h_l}$ in the form 
\beq
d\sigma^{h_p h_l}= d \sigma_0 + h_p d\sigma_{\rm SSA} + h_l d\sigma_{\rm BSA} + h_p h_l d\sigma_{\rm DSA}, 
\eeq
where the last three terms represent  single spin asymmetry (SSA), beam spin asymmetry (BSA) and double spin asymmetry (DSA). The previous works \cite{Ji:2016jgn,Hatta:2016aoc} looked for  signals of OAM in SSA, while our main focus here is DSA  which can be isolated by forming the linear combination 
\beq
d\sigma_{\rm DSA}= \frac{1}{4}(d\sigma^{++} - d\sigma^{+-} -d\sigma^{-+} + d\sigma^{--}).
\eeq

The incoming proton has momentum $p^\mu\approx ((1+\xi)P^+,0,0_\perp)$, where $\xi$ is the skewness variable, and scatters elastically with momentum transfer $\Delta^\mu=(-2\xi P^+,0,\Delta_\perp)$ with $\Delta_\perp=|\Delta_\perp|(\cos \phi_{\Delta_\perp},\sin \phi_{\Delta_\perp})$. The virtual photon's momentum is taken in the form  
\beq
q^\mu = \left(-\frac{Q}{\sqrt{2}}, \frac{Q}{\sqrt{2}}, 0_\perp\right), \label{qmu}
\eeq
with $q^2=-Q^2$. 
In the present frame, the incoming lepton momentum is given by 
\beq
l^\mu = \left( \frac{Q(1-y)}{\sqrt{2}y}, \frac{Q}{\sqrt{2}y}, \frac{Q\sqrt{1-y}}{y}n_\perp\right),
\eeq 
where $y=p\cdot q/p\cdot l$ is the usual variable in DIS and $n_\perp = (\cos \phi_{l_\perp},\sin \phi_{l_\perp})$ is a unit vector in the transverse plane. The lepton mass is neglected so that the lepton spin vector is $s^\mu \approx h_l l^\mu$.   The center-of-mass energies of the $ep$ and $\gamma^*p$ collisions are denoted by $s_{ep}=(p+l)^2$ and $W^2=(p+q)^2 = M^2-Q^2+ys_{ep}$, respectively. 

 The signal of OAM proposed in \cite{Bhattacharya:2022vvo} is an azimuthal correlation of the form 
\beq
d\sigma_{\rm DSA} \propto \Delta_\perp\cdot l_\perp = |\Delta_\perp||l_\perp|\cos (\phi_{\Delta_\perp}-\phi_{l_\perp}), \label{ang}
\eeq 
measured in coincidence with the production of two jets (dijet) whose azimuthal angle is integrated over. 
We assume $\Delta_\perp$ is small and keep only linear terms in $\Delta_\perp$. Under this assumption, we parameterize the momenta of the final state quark and antiquark  as 
\beq
\begin{split}
& q_1^\mu  = \frac{q_\perp^2+\mu^2}{4\xi\bar{z}} n^\mu +  \frac{2\xi\bar{z} q_\perp^2}{q_\perp^2+\mu^2}\left(1-2z\frac{q_\perp \cdot \Delta_\perp}{q_\perp^2}\right) P^\mu + \delta^\mu_i(q_\perp^i -z\Delta^i_\perp), \label{q12}\\
&q_2^\mu  =  \frac{q_\perp^2+\mu^2}{4\xi z} n^\mu +  \frac{2\xi z  q_\perp^2}{q_\perp^2+\mu^2}\left(1+2\bar{z}\frac{q_\perp \cdot \Delta_\perp}{q_\perp^2}\right) P^\mu +\delta^\mu_i(-q_\perp^i -\bar{z}\Delta^i_\perp)  ,
\end{split} 
\eeq
where  $\mu^2\equiv z\bar{z}Q^2$.  $n^\mu = \delta^\mu_-/P^+$ and $P^\mu=\delta^\mu_+P^+$ with $n\cdot P=1$ are orthogonal light-like vectors.  
The condition  $q_1^2=q_2^2=0$ is satisfied to linear order in $\Delta_\perp$. We have introduced the common variables 
 $z=P\cdot q_1/P\cdot q=q_1^-/q^-$ and $\bar{z}\equiv 1-z=q_2^-/q^-$ denoting the longitudinal momentum fractions of the virtual photon carried by the quark   and the antiquark, respectively. 
The skewness variable is related to $q_\perp$ as 
\begin{align}
    \xi (q^{2}_\perp)  & = \dfrac{q^{2}_\perp+ \mu^2}{-q^{2}_\perp + \mu^2+2z\overline{z}W^{2}} . \label{xiq}
\end{align}
At high center-of-mass energy ($W\sim 100$ GeV at the EIC), typically $\xi \ll 1$. 

 The spin-dependent part of the lepton tensor is 
\beq
L^{\mu\nu}=\sum_{s'}\bar{u}(ls)\gamma^\mu u(l's')\bar{u}(l's')\gamma^\nu u(ls) \sim -2i\epsilon^{\rho \tau\mu\nu}s_\rho q_\tau =-2ih_l \epsilon^{\rho \tau\mu\nu}l_\rho q_\tau .
\eeq
Since the observable (\ref{ang}) depends on $\phi_{l_\perp}$, the index $\rho$ has to be transverse. On the other hand, the index  $\tau$ is longitudinal (see  (\ref{qmu})). This means  that either $\mu$ or $\nu$ must be transverse and the other  longitudinal. In other words, we are after the interference effect between   amplitudes with longitudinally  polarized and transversely  polarized virtual photons. At the same time, the cross section also encodes  interference  between twist-two and twist-three amplitudes because the dependence on $\Delta_\perp$ shows up only at the twist-three level.  Let us thus denote the longitudinal ($L$) and transverse ($T$) amplitudes as  
\beq
A_L\equiv \epsilon_\mu^LA^\mu= A_L^2+A_L^3, \qquad A_{T\lambda} \equiv \epsilon^{ i}_{T\lambda} A^i = \epsilon_{T\lambda}^{i} (A_{T}^{2i}+A_{T}^{3i}),
\eeq
where $\lambda=1,2$ is a label for the two transverse polarization vectors and 
\beq
\epsilon_\mu^L =\frac{1}{Q}\left(q_\mu+\frac{Q^2}{p\cdot q}p_\mu\right), 
\eeq
is the longitudinal photon polarization vector. The term proportional to $q^\mu$ can be omitted due to gauge invariance.\footnote{Strictly speaking, the QED Ward identity does not hold after a twist expansion. However, the corrections either do not affect the $\cos (\phi_{\Delta_\perp}-\phi_{l_\perp})$ dependence or are of higher order in $\Delta_\perp,k_\perp$. 
}   
The superscripts 2 and 3 denote the twist at which the amplitude is evaluated. The relevant part of the cross section is  then given by 
\beq
L^{\mu\nu}A^*_\mu A_\nu &\sim&  - \sum_{\lambda=1,2} L^{\mu\nu} (\epsilon^L_\mu\epsilon_\nu^{T\lambda*}A^*_LA^{\lambda}_T+ \epsilon^{T\lambda}_\mu\epsilon_\nu^{L}A_T^{\lambda*}A_L)\nn
&=& 2ih_l|l_\perp|\epsilon_{ij}  (A_L^{2*}A^{3i}_{T} +A_L^{3*}A^{2i}_{T}  -A^{2i*}_{T}A^3_L - A^{3i*}_{T}A^2_L ) n_{\perp }^j\nn
 &=&4h_l|l_\perp|\epsilon_{ij} {\rm Re} (iA_L^{2*}A^{3i}_{T}   -iA^{2i*}_{T}A^3_L  ) n_{\perp }^j . \label{interf}
\eeq
Dirac traces involving the incoming and outgoing nucleon spinors, as well as the produced $q\bar{q}$ spinors are understood.  The cosine angular correlation (\ref{ang}) originates from the following structure 
\beq
 {\rm Re} (iA_L^{2*}A^{3i}_{T}   -iA^{2i*}_{T}A^3_L  ) \sim h_p\epsilon^{ik}\Delta^k_\perp. \label{fol}
\eeq
As observed in \cite{Bhattacharya:2022vvo}, there are two different but  equally important (at least parametrically) contributions to (\ref{fol}). A factor of $\Delta_\perp$ can arise either from the soft (GTMD) part  or the hard scattering part. 
The former is directly related to the gluon OAM which we are mainly interested in, and  the latter is the so-called kinematical twist-three effect which involves only twist-two GPDs (hence unrelated to GTMDs).  We shall discuss these contributions in the next subsection. 

\subsection{ Outline of the calculation}
\begin{figure}[htbp]
  \includegraphics[width=0.3\linewidth]{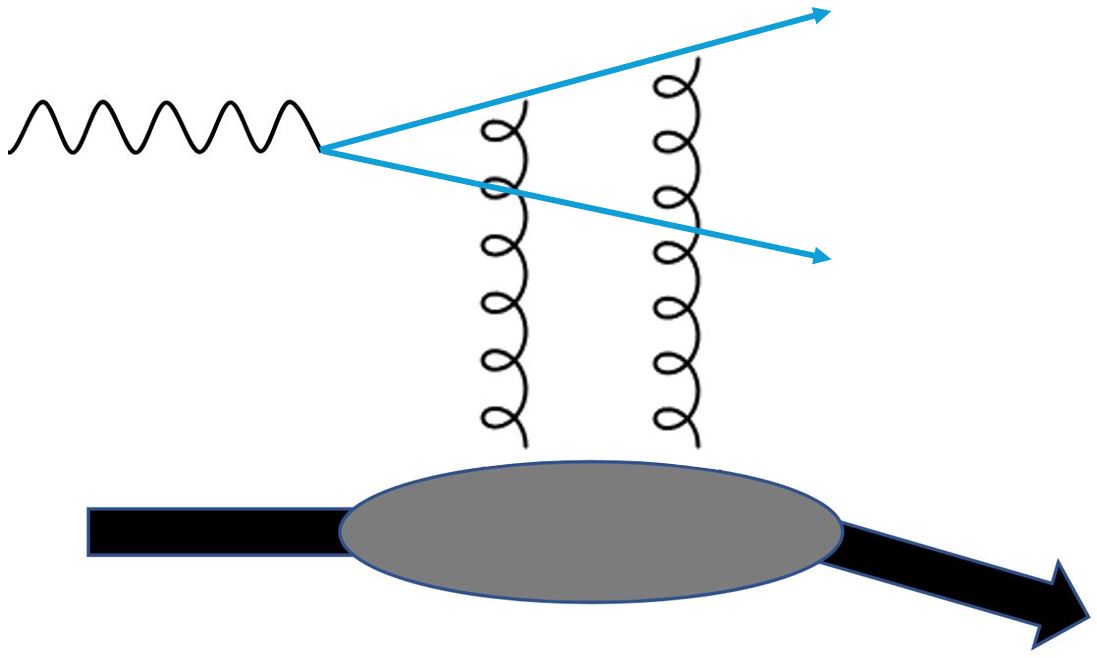}
   \includegraphics[width=0.3\linewidth]{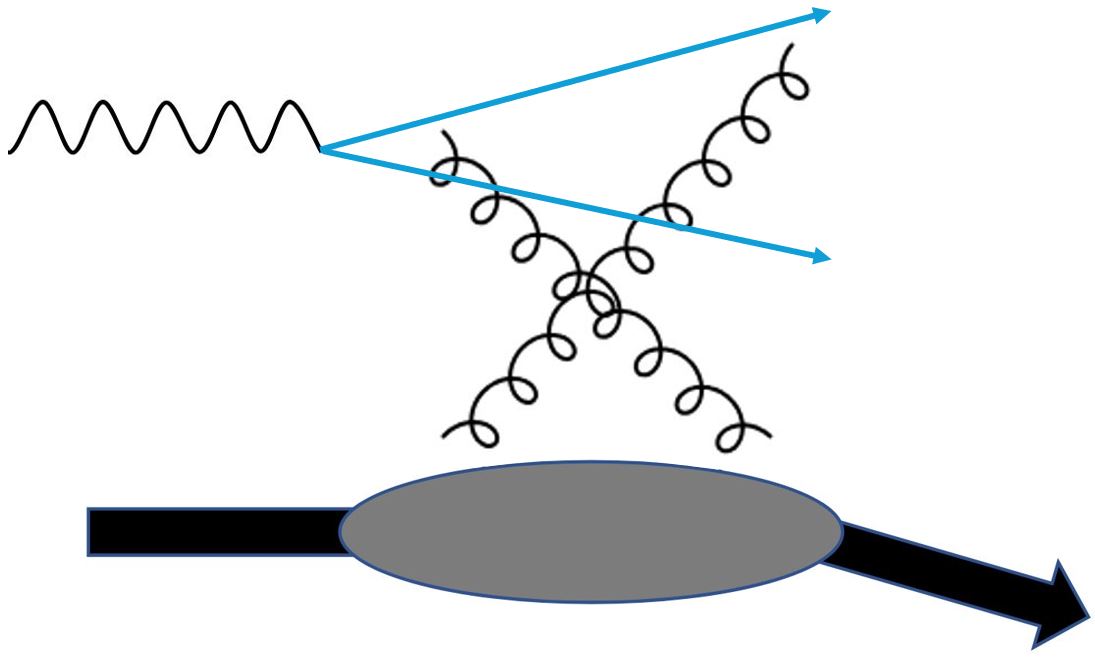}
   \includegraphics[width=0.3\linewidth]{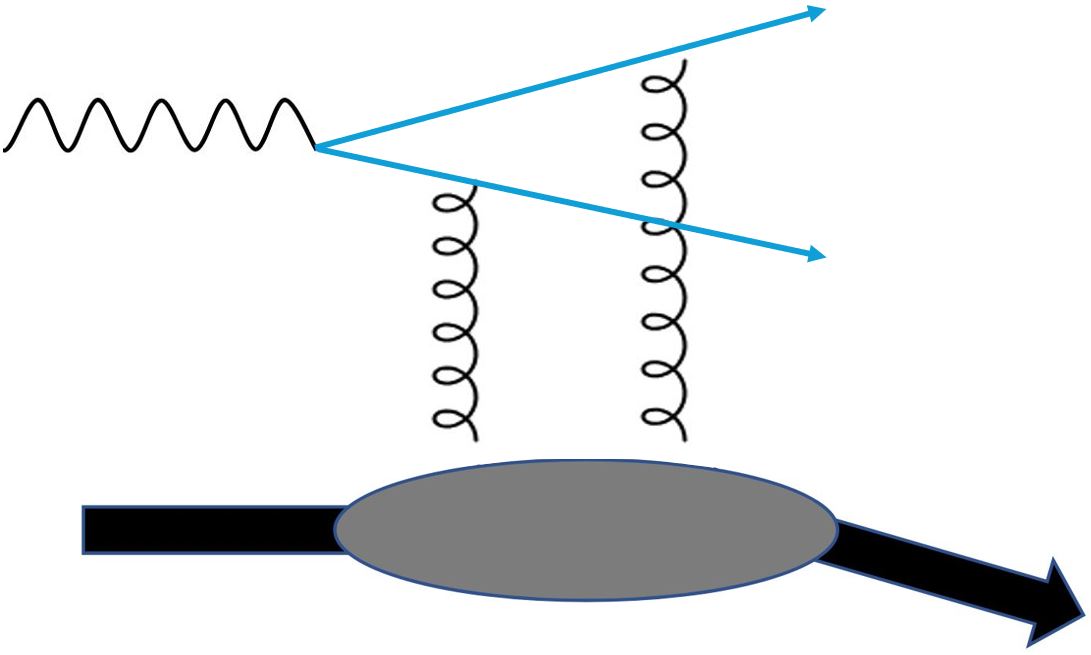}
\caption{Presented are the leading-order Feynman diagrams depicting exclusive dijet production initiated by gluons. The complementary set of three diagrams, with the quark lines exchanged, is implicit.}
\label{fig:gluons}
\end{figure} 
In the two-gluon exchange approximation, the hadronic part of the scattering amplitude can be written as 
\beq
i A_{L/T}  =\frac{-ig^2e_{em}e_qC_F}{\left(N_{c}^{2}-1\right)P^{+}}\int\frac{dz^{-}d^{2}z_\perp}{\left(2\pi\right)^{3}}\int\frac{dx}{x^{2}-\xi^{2}}\int d^{2}k_\perp e^{ixP^{+}z^{-}-ik_\perp\cdot z_\perp} D^{L/T}_{\mu\nu}\left\langle  p^{\prime}\left|F^{+\mu}\left(-\frac{z}{2}\right)F^{+\nu}\left(\frac{z}{2}\right)\right|p\right\rangle . \label{master}
\eeq
Here and below, $\frac{1}{x^2-\xi^2}$ is a shorthand notation for 
\beq
\frac{1}{(x-\xi+i\epsilon)(x+\xi-i\epsilon)}.
\eeq
The hard part $D_{\mu\nu}$ consists of 
six diagrams as shown in Fig.~\ref{fig:gluons}. 
Explicitly, 
\beq
D^{L/T}_{\mu\nu}&=&\epsilon^\rho_{L/T}\bar{u}(q_1) \Biggl[ \gamma_\nu \frac{1}{\Slash q_1+\Slash k+\Slash \Delta/2}\gamma_\mu \frac{1}{\Slash q_1+\Slash \Delta} \gamma_\rho+ \gamma_\rho\frac{1}{\Slash q_2+\Slash \Delta}\gamma_\mu \frac{1}{\Slash q_2+\Slash k + \Slash \Delta/2}\gamma_\nu \nn 
&&\qquad \quad + \gamma_\mu \frac{1}{\Slash q_1-\Slash k +\Slash \Delta/2}\gamma_\nu \frac{1}{\Slash q_1+\Slash \Delta}\gamma_\rho + \gamma_\rho \frac{1}{\Slash q_2+\Slash \Delta}\gamma_\nu \frac{1}{\Slash q_2-\Slash k + \Slash \Delta/2} \gamma_\mu \nn 
&& \qquad \quad  -\gamma_\mu \frac{1}{\Slash q_1-\Slash k+\Slash \Delta/2}\gamma_\rho \frac{1}{\Slash q_2+\Slash k+\Slash \Delta/2}\gamma_\nu - \gamma_\nu \frac{1}{\Slash q_1+\Slash k+\Slash \Delta/2}\gamma_\rho \frac{1}{\Slash q_2-\Slash k+\Slash \Delta/2}\gamma_\mu \Biggr]v(q_2) ,\label{hard}
\eeq
where $k^\mu=(xP^+,0,k_\perp)$. 
One then projects onto the symmetric (unpolarized) and antisymmetric (polarized) parts in the gluon transverse indices 
\beq
D_{\mu\nu}\langle F^{+\mu}F^{+\nu} \rangle&=& \frac{1}{2}g_{\mu\nu}^\perp  D^{\mu\nu} \langle g_{\alpha\beta}^\perp F^{+\alpha}F^{+\beta}\rangle + \frac{1}{2}\epsilon^\perp_{\mu\nu}D^{\mu\nu}\langle \epsilon^\perp_{\alpha\beta}F^{+\alpha}F^{+\beta}\rangle\nn
&\sim& \frac{1}{2}\delta_{ij}D_{ij} xf_g + \frac{1}{2}\epsilon_{ij}D_{ij} ix\tilde{f}_g,
\label{symasym}
\eeq
where $g_{\mu\nu}^\perp = g_{\mu\nu}-P_\mu n_\nu -P_\nu n_\mu$ and $\epsilon_{\mu\nu}^\perp = \epsilon_{\mu\nu\alpha\beta}P^\alpha n^\beta$.

Our basic strategy is to expand $D_{ij}$ in powers of $k_\perp$ and $\Delta_\perp$ to linear order 
\beq
D_{ij}(k_\perp,\Delta_\perp)\approx D_{ij}(0_\perp,0_\perp) + \left.\frac{\partial D_{ij}(k_\perp,0_\perp)}{\partial k_\perp^l} \right|_{k=0}k_\perp^l + \left.\frac{\partial D_{ij}(0_\perp,\Delta_\perp)}{\partial \Delta_\perp^l}\right|_{\Delta=0} \Delta_\perp^l. \label{linear}
\eeq
To do so, we need to expand each building block of (\ref{hard}). For example, the first denominator in (\ref{hard}) becomes 
\beq
\frac{1}{(q_1+k+\Delta/2)^2}&\approx& \frac{1}{\frac{q_\perp^2+\mu^2}{2\xi\bar{z}}(x-\xi)-2q_\perp \cdot k_\perp-q_\perp \cdot \Delta_\perp} \nn
&\approx& \frac{2\xi\bar{z}}{(q_\perp^2+\mu^2)(x-\xi)} + \frac{(2\xi\bar{z})^2}{(q_\perp^2+\mu^2)^2(x-\xi)^2}\left(2q_\perp\cdot k_\perp +q_\perp \cdot \Delta_\perp\right).\label{expand}
\eeq
When (\ref{linear}) is substituted into (\ref{symasym}), the integral $\int d^2k_\perp k_\perp^l xf_g \sim \epsilon^{lm}\Delta_\perp^m L_g$ establishes a connection to the gluon OAM (\ref{lgx}). On the other hand,  the first and last terms in (\ref{linear}) are independent of $k_\perp$, so these terms lead  to the twist-two gluon GPDs $\int d^2k_\perp xf_g\sim H_g, E_g$ and $\int d^2k_\perp x\tilde{f}_g\sim \tilde{H}_g, \tilde{E}_g$. Further squaring the amplitude as in (\ref{interf}), we encounter a number of interference terms. We have considered  all the possibilities and found that the angular dependence (\ref{ang}) can come from the following   sources
\beq
H_g\otimes  \int d^2k_\perp k^l_\perp  xf_g, \qquad H_g \otimes \tilde{H}_g, \qquad \tilde{H}_g\otimes\int d^2k_\perp k^l_\perp x\tilde{f}_g. \label{few}
\eeq
 The first term involves the gluon OAM and the second term represents the  kinematical twist-three effect mentioned above. 
In  \cite{Bhattacharya:2022vvo}, we neglected the third term assuming that, at high  energy $\xi\ll 1$,  the product of the polarized GPD and the polarized GTMD would be subleading. In the following, we shall also calculate the third term to check the validity of this assumption.

First, consider   the twist-two ($k_\perp=\Delta_\perp=0$) term in (\ref{linear}). Applying the symmetric projection $\delta_{ij} D_{ij}$ (denoted by a superscript $\delta$), one finds  \cite{Braun:2005rg}\footnote{ The following formulas which can be derived from the Dirac equations $\bar{u}(q_1)\Slash q_1=\Slash q_2v(q_2)=0$ are useful when relating seemingly different expressions in the literature
\beq
\bar{u}(q_1)\left[ (z-\bar{z})q_\perp \cdot \epsilon_\perp \gamma^- - i\epsilon_\perp \times q_\perp \gamma_5\gamma^- + \epsilon_\perp\cdot \gamma_\perp \frac{q_\perp^2+\mu^2}{2\xi P^+} \right] v(q_2) \approx 0,
\eeq
\beq
\bar{u}(q_1)\left[ iq_\perp \cdot \epsilon_\perp \gamma_5\gamma^- +(z-\bar{z})\epsilon_\perp \times q_\perp\gamma^- +\epsilon_\perp \times \gamma_\perp \frac{q_\perp^2+\mu^2}{2\xi P^+} \right] v(q_2) \approx 0,
\eeq
where $\approx$ means neglecting terms of order $\Delta_\perp$.   }
\begin{equation}
\begin{split}
& iA^{2\delta}_T=  - \dfrac{i g^{2} e_{em} e_q}{N_c} \frac{\bar{u}(q_1) \epsilon_\perp\cdot \gamma_\perp v(q_2)}{q_\perp^2+\mu^2} \int dx \frac{1}{x^2-\xi^2} \left(1+\frac{2\xi^2(1-2\beta)}{x^2-\xi^2}\right) \int d^2k_\perp xf_g(x,\xi,k_\perp,\Delta_\perp), \label{at}
\\
& iA^{2\delta}_L =   \dfrac{i g^{2} e_{em} e_q}{N_c} \frac{Q}{(q_\perp^2+\mu^2)q^-} \bar{u}(q_1)\gamma^-v(q_2) \int dx \frac{1}{x^2-\xi^2} \left(1+\frac{4\xi^2(1-\beta)}{x^2-\xi^2}\right)\int d^2k_\perp xf_g(x,\xi,k_\perp,\Delta_\perp), 
\end{split}
\end{equation}
where 
\beq
\beta \equiv \frac{\mu^2}{q_\perp^2+\mu^2}.\label{beta}
\eeq
As we shall discuss later, $\beta$ is related to the invariant mass of the dijet. It is actually a commonly used variable in the context of diffractive DIS. The following relation is useful throughout the calculation
\beq
 P^+q^-=  \frac{q_\perp^2+\mu^2}{4\xi z\bar{z}}= \frac{q_\perp^2}{4\xi z\bar{z}(1-\beta)}.
\eeq
The antisymmetric contraction (denoted by a superscript $\epsilon$) gives 
\begin{equation}
\begin{split}
&iA_T^{2\epsilon} =   \dfrac{i g^{2} e_{em} e_q}{N_c} \frac{\bar{u}(q_1) \epsilon_\perp  \times \gamma_\perp 
 v(q_2)}{q_\perp^2+\mu^2} \int dx \frac{2\xi x}{(x^2-\xi^2)^2} \int d^2k_\perp x\tilde{f}_g(x,\xi,k_\perp,\Delta_\perp), \\
& iA_L^{2\epsilon}=0. \label{it}
\end{split}
\end{equation}
The $k_\perp$-integrals in  (\ref{at}) and (\ref{it}) convert the gluon GTMDs into    the  gluon GPDs $H_g, E_g,\tilde{H}_g,\tilde{E}_g$. Further integration over $x$ yields the following moments of the GPDs 
\begin{equation}
\begin{split}
& {\cal H}^{(1)}_g(\xi)=\int^1_{-1} dx \frac{ H_g(x,\xi)}{x^2-\xi^2} , \qquad {\cal E}^{(1)}_g(\xi)=\int^1_{-1} dx \frac{ E_g(x,\xi)}{x^2-\xi^2}, \\ 
& {\cal H}^{(2)}_g(\xi)=\int^1_{-1} dx  \frac{\xi^2 H_g(x,\xi) }{(x^2-\xi^2)^2}, \qquad {\cal E}^{(2)}_g(\xi)=\int^1_{-1} dx  \frac{\xi^2 E_g(x,\xi) }{(x^2-\xi^2)^2}, \\
& \tilde{{\cal H}}^{(2)}_g(\xi)=\int^1_{-1} dx  \frac{x \tilde{H}_g(x,\xi) }{(x^2-\xi^2)^2}, \qquad \tilde{{\cal E}}^{(2)}_g(\xi)=\int^1_{-1} dx  \frac{x \tilde{E}_g(x,\xi) }{(x^2-\xi^2)^2}.
\label{h2}
\end{split}
\end{equation}
These integrals are well defined as long as the GPDs  and their first derivatives are continuous at $x=\pm \xi$, which is usually considered to be the case. Squaring the amplitudes (\ref{at}), one obtains the cross section of unpolarized dijet production \cite{Braun:2005rg}. 

We now turn to the  twist-three contributions  that give rise to the asymmetry (\ref{ang}). The first entry in (\ref{few}) related to the OAM originates from the second term in (\ref{linear}). 
The calculation of the amplitude has been done in \cite{Ji:2016jgn} in the context of SSA, and the same results are relevant to DSA:
\begin{equation}
\begin{split}
iA^{3\delta}_{T} & = - \dfrac{i g^{2} e_{em} e_q}{N_c} \dfrac{4\xi(\overline{z}-z)}{(q^{2}_\perp + \mu^{2})^{2}}  \bar{u}(q_1) \epsilon_\perp \cdot \gamma_\perp v(q_2) \\
& \qquad \qquad \times\int dx \dfrac{x}{ (x^2 - \xi^2 )^{2}}  \bigg ( 1 + \dfrac{4\xi^2(1-2\beta)}{x^2-\xi^2}\bigg ) \int d^{2}k_\perp q_\perp \cdot k_\perp \, x f_{g}(x,\xi,k_\perp,\Delta_\perp) \\[0.2cm]
& \quad - \dfrac{i g^{2} e_{em} e_q}{N_c} \dfrac{2\xi }{(q^{2}_\perp + \mu^{2})q^-}  \bar{u}(q_1) \gamma^- v(q_2)\int dx \dfrac{x}{(x^2 - \xi^2 )^{2} }\int d^{2}k_\perp \epsilon_\perp \cdot k_\perp \, x f_{g}(x,\xi,k_\perp,\Delta_\perp), \label{a3t1} \\[0.3cm] 
iA^{3\delta}_{L} & = \dfrac{i g^{2} e_{em} e_q}{N_c} \dfrac{4 \xi(\overline{z}-z)Q }{(q^{2}_\perp + \mu^{2})^2q^-}  \bar{u}(q_1) \gamma^- v(q_2) \\
& \qquad \qquad \times \int dx \dfrac{x}{(x^2 - \xi^2)^2 } 
\bigg ( 1 + \dfrac{8 \xi^{2}(1-\beta)}{x^2-\xi^2 } \bigg )  \int d^{2}k_\perp \, q_\perp \cdot k_\perp \, x f_g (x,\xi, k_\perp, \Delta_\perp ) . 
\end{split}
\end{equation}
For this contribution, the factor $\Delta_\perp$ necessary for the angular dependence (\ref{ang}) emerges from the Dirac trace over the nucleon spinors. Indeed, taking the interference effect between (\ref{at}) and  (\ref{a3t1}) and summing over the outgoing nucleon spins while fixing the spin of the incoming nucleon, we find 
\begin{align}
 &\frac{1}{4P^+M} {\rm tr}\left[ \left(H_g\gamma^+ -\frac{i\sigma^{+\nu}\widetilde{\Delta}_\nu}{2M}E_g\right) \left(\Slash p^{\prime}+M\right)\left(F^g_{1,1}+i\frac{\sigma^{j+}}{P^+}(\widetilde{k}_\perp^j F^g_{1,2}+\widetilde{\Delta}_\perp^j F^g_{1,3})+ i\frac{\sigma^{ij}\widetilde{k}_\perp^i\widetilde{\Delta}_\perp^j}{M^2}F^g_{1,4}\right)\left(\Slash p+M\right)\frac{\Slash s\gamma_{5}}{2M}\right] \nn
& =ih_{p}\frac{1+\xi}{2M^2}\epsilon^{ij}k_\perp^{i}\Delta_\perp^{j}\left[2\left(H_{g}-\frac{\xi^{2}}{1-\xi^{2}}E_{g}\right)F_{1,4}^{g}-E_{g}F_{1,2}^{g}\right].\label{trac} 
\end{align}
 As indicated by the notations $\widetilde{k}_\perp,\widetilde{\Delta}_\perp$, this can be conveniently evaluated in the symmetric frame where the factor of $1+\xi$ comes from  $\tilde{\Delta}_\perp =(1+\xi)\Delta_\perp$.
 The product of $k_\perp^i F_{1,4}^g$ from (\ref{trac}) and    $k_\perp$ from (\ref{a3t1}) leads to  the moment   (\ref{lgx}) after the $k_\perp$ integration.   From the $k_\perp^i F_{1,2}^g$ term in (\ref{trac}), we get another moment
\beq
xO(x,\xi)\equiv \int d^2k_\perp \frac{k_\perp^2}{M^2}F^g_{1,2}(x,\xi). \label{oint}
\eeq 
The notation $O$ is a reminder that the imaginary part of $F_{1,2}^g$ is related to the so-called spin-dependent Odderon \cite{Zhou:2013gsa,Szymanowski:2016mbq}.  Its second moment at $\xi=0$ is proportional to the $C$-odd three-gluon correlator $\langle P |d_{abc}F_a^{+i}F_b^{+j}F_c^{+i}|P\rangle$ \cite{Ji:1992eu} relevant to transverse single spin asymmetry \cite{Zhou:2013gsa}.\footnote{We note that if one chooses the Weisz\"acker-Williams gluon distribution in (\ref{wwdipole}), the moment (\ref{oint}) is instead related to the $C$-even three-gluon correlator $\langle P |if_{abc}F_a^{+i}F_b^{+j}F_c^{+i}|P\rangle$.}  The real part of $F_{1,2}^g$ is proportional to $\xi$ (at $\Delta_\perp=0$, see (\ref{real12})) so it is suppressed when $\xi \ll 1$.\footnote{Recently it has been shown in \cite{Agrawal:2023cdf} that the real part of $F_{1,2}^g$  has a singular small-$x$ behavior $\sim 1/x^c$ with $c>0$ unlike the imaginary part.  It may become relevant once higher order effects in $\Delta_\perp$ are considered.}

The subsequent integration over $x$ in (\ref{a3t1}) yields  the following moments
\beq
{\cal L}^{(2)}_g(\xi) &=&\int^1_{-1} dx \frac{x^2L_g(x,\xi)}{(x^2-\xi^2)^2 } ,\qquad 
{\cal O}^{(2)}(\xi) =\int^1_{-1} dx \frac{x^2O(x,\xi)}{(x^2-\xi^2)^2 }, \label{l1}  \\
{\cal L}^{(3)}_g(\xi) &=&\int^1_{-1} dx \frac{\xi^2x^2L_g(x,\xi)}{(x^2-\xi^2)^3 } ,\qquad 
{\cal O}^{(3)}(\xi) =\int^1_{-1} dx \frac{\xi^2 x^2O(x,\xi)}{(x^2-\xi^2)^3 }. \label{lg3}
\eeq
We immediately notice triple poles at $x=\pm \xi$ which were absent in the twist-two amplitudes   (\ref{h2}).  In order for the $x$-integrals in (\ref{lg3}) to be well-defined, the second derivative of $L_g(x,\xi)$ and $O(x,\xi)$ must be continuous at $x=\pm\xi$. It is known that such a condition can be violated for gluonic GPDs, and this in fact signals  the violation of collinear factorization. We postpone  this issue until the next subsection where the problem becomes sharper. In the meantime, 
let us just point  out that all the terms in (\ref{a3t1}) which contain triple poles can be eliminated by setting $z=\bar{z}=\frac{1}{2}$, that is, when the produced jets are symmetric.

Next, the second entry in (\ref{few}) comes from  the last term in (\ref{linear}) obtained by  setting $k_\perp=0$ and keeping one factor of $\Delta_\perp$ in the hard part.    
Accordingly,  in the soft part one can set $\Delta_\perp=0$ and integrate over $k_\perp$, after which GTMDs reduce to GPDs in both the amplitude and the complex-conjugate amplitude.
The structure (\ref{fol}) then arises from interference  between the unpolarized and polarized amplitudes.  The asymmetry is therefore proportional to   
\begin{align}
 & \frac{1}{4(P^+)^2}{\rm tr}\left[\left(H_g\gamma^+ -\frac{i\sigma^{+\nu}\Delta_\nu}{2M}E_g\right)\left(\Slash p^{\prime}+M\right)  \left(i\gamma_{5}\gamma^{+}\widetilde{H}_{g}-i\frac{\gamma_{5}\Delta^{+}}{2M}\widetilde{E}_{g}\right) \left(\Slash p+M\right)\frac{\Slash s\gamma_{5}}{2M}\right]\nn
 & =ih_{p}\left(1-\xi^{2}\right)\left(H_{g}-\frac{\xi^{2}}{1-\xi^{2}}E_{g}\right)\left(\widetilde{H}_{g}-\frac{\xi^{2}}{1-\xi^{2}}\widetilde{E}_{g}\right)
 \label{trac2}.
\end{align}

The calculation of the hard part is straightforward but quite cumbersome, since one has to consider many cases: longitudinal/transverse photons; twist-two/twist-three amplitudes;  symmetric/antisymmetric projections (\ref{symasym}). Moreover, in addition to the explicit $\Delta_\perp$-dependence coming from the expansion of the propagators, cf., (\ref{expand}),  the factor of $\Delta_\perp$ can also come from the Dirac trace over the final state quark spinors $\bar{u}(q_1),v(q_2)$ since $q_{1,2}$ depend on $\Delta_\perp$, see (\ref{q12}). Therefore, in practice it is more advantageous to expand the squared  amplitude in $\Delta_\perp$ instead of the amplitude itself. Schematically,  (\ref{interf}) should be generalized as 
\beq
{\rm Re}\left[iA_L^{2\delta *}A_T^{3\epsilon i} +iA_L^{3\delta *}A_T^{2\epsilon i}- iA_T^{2\delta i*}A_L^{3\epsilon}-iA_T^{3\delta i*}A_L^{2\epsilon} + i(A_L^{\delta*} A_T^{\epsilon i})^3-i(A_T^{\delta i*}A_L^{\epsilon})^3 \right]. 
\eeq
 In the last two terms,   the factor of $\Delta_\perp$ comes from the squared hard amplitude including the final state  $q\bar{q}$ spinors.  
In  \cite{Bhattacharya:2022vvo} we have performed this calculation analytically for the special case $z=\frac{1}{2}$. For generic values of $z$, we have developed a  Mathematica program to facilitate the calculation.

Finally, we turn to the third entry in (\ref{few}). This can be calculated in the same way as the OAM contribution except that $\delta_{ij}\to \epsilon_{ij}$ everywhere. 
 The twist-three hard amplitudes read 
\begin{equation}
\begin{split}
iA^{3\epsilon}_{T} & = - \dfrac{i g^{2} e_{em} e_q}{N_c} \int \frac{dx}{(x^2-\xi^2)^2} \Biggl[ \left\{ \bar{u}(q_1) \epsilon_\perp \times \gamma_\perp v(q_2) (z-\bar{z})\left(3+\frac{4\xi^2}{x^2-\xi^2}\right) +  \bar{u}(q_1)\gamma^-v(q_2)\frac{2q_\perp \times \epsilon_\perp}{q^-}\right\} \\
& \qquad \qquad \qquad \qquad \qquad  \times \frac{4\xi^2}{(q_\perp^2+\mu^2)^2} \int d^{2}k_\perp q_\perp \cdot k_\perp \, ix \tilde{f}_{g}(x,\xi,k_\perp,\Delta_\perp) \\[0.2cm]
& \qquad \qquad \qquad  \quad +\bar{u}(q_1)\gamma^-v(q_2)\frac{2\xi^2}{(q_\perp^2+\mu^2) q^-}  \int d^{2}k_\perp \epsilon_\perp \times k_\perp \, ix \tilde{f}_{g}(x,\xi,k_\perp,\Delta_\perp) \Biggr], \label{a3t2} \\[0.3cm] 
iA^{3\epsilon}_{L} & = -\dfrac{i g^{2} e_{em} e_q}{N_c}    \frac{4\xi^2 Q}{q^-(q_\perp^2+\mu^2)^2}\bar{u}(q_1) \gamma_5\gamma^- v(q_2)  \int \dfrac{dx}{(x^2 - \xi^2)^2 } 
 \int d^{2}k_\perp \, q_\perp \cdot k_\perp \, x \tilde{f}_g (x,\xi, k_\perp, \Delta_\perp ) . 
\end{split}
\end{equation}
In the soft part, the nucleon spinor sum gives 
\begin{align}
 & \frac{-i}{4P^+M} {\rm tr}\left[\left(\tilde{H}_g\gamma_5\gamma^+ +\tilde{E}_g \frac{\gamma_5\Delta^+}{2M} \right)\left(\Slash p^{\prime}+M\right)  \left(\frac{\epsilon_{ij}\widetilde{k}_\perp^i \widetilde{\Delta}_\perp^j}{M^2}G^g_{1,1}-\frac{\sigma^{i+}\gamma_5}{P^+}(\widetilde{k}_\perp^i G^g_{1,2}+\widetilde{\Delta}_\perp^i G^g_{1,3})-\sigma^{+-}\gamma_5 G^g_{1,4} \right) \left(\Slash p+M\right)\frac{\Slash s\gamma_{5}}{2M}\right]\nn
 & =ih_p\frac{1+\xi}{2M^2}\left(2\left(\tilde{H}_g -\frac{\xi^2}{1-\xi^2}\tilde{E}_g\right)G^g_{1,1}-\xi \tilde{E}_gG^g_{1,2}\right) \epsilon_{ij}k_\perp^i \Delta^j_\perp 
 \label{trac3}.
\end{align}
The $k_\perp$ integrals then lead to the moments
\beq
x C_g(x,\xi)\equiv \int d^2 k_\perp\frac{k_\perp^2}{M^2}G_{1,1}^g(x,\xi), \qquad xC'_g(x,\xi)\equiv -\int d^2 k_\perp\frac{k_\perp^2}{M^2}G_{1,2}^g(x,\xi).  \label{orbit}
\eeq
With these definitions, $C_g(x,\xi)$ and $C'_g(x,\xi)$ are odd functions in $x$. [Note that $H_g(x,\xi), x\tilde{H}_g(x,\xi), L_g(x,\xi)$ and $O(x,\xi)$ are all even functions in $x$.] Similarly to the quark case \cite{Lorce:2011kd}, we interpret $C_g(x,\xi=0)$ as the gluons' spin-orbit correlation. On the other hand, $C'_g(x,\xi=0)$ is related to another $C$-odd three-gluon correlator $\langle d_{abc} F_a^{+i}F_b^{+j}\tilde{F}_c^{+i}\rangle$.  Very little is known about these distributions  except that  
they appear in specific exclusive reactions  \cite{Bhattacharya:2018lgm,Boussarie:2018zwg}.

Let us provide quick insights into the characteristics of $C_g(x)$ which allows for a quantitative estimate sufficient for the present purpose, leaving a more detailed analysis for future work. First, despite being associated with the polarized gluon operator $\tilde{F}^{+\mu} F^+_{\ \mu}$, $C_g(x)$ has little to do with the parent hadron spin and exists even for  spinless hadrons. Second,  
just like $L_g(x)$ (see (\ref{www})), using the equations of motion one can express $C_g(x)$ as the sum of the Wandzura-Wilczek (WW) part and the genuine twist-three part. Adapting the method in \cite{Hatta:2012cs}, we have computed the WW part 
\beq
 C_g(x) &=&  x\int_x^{\epsilon(x)}\frac{dx'}{x'^2}\tilde{H}_g(x') -2x \int_x^{\epsilon(x)} \frac{dx'}{x'^2}G(x')+\cdots. \label{g11}
\eeq
(\ref{g11}) allows us to infer the small-$x$ behavior of $C_g(x)$. Clearly, the second term dominates at small-$x$ because,  roughly, $\tilde{H}_g(x) =x\Delta G(x) \sim x^2 G(x)$. Assuming the power-law (\ref{bfkl}), we find 
\beq
C_g(x) \approx -\frac{2}{2+c} G(x) \approx -G(x),  \label{cgx}
\eeq
where the last expression follows if the exponent is perturbatively small $c\propto \alpha_s \ll 1$.  Since $G(x)$ is positive at small-$x$, $C_g(x)$ is negative, meaning that  the helicity and OAM of individual gluons are anti-aligned (have opposite signs) \cite{Lorce:2011kd} at small-$x$, in accordance  with the theoretical prediction (\ref{lww}). (\ref{cgx}) also shows that $C_g(x)$ grows strongly in the small-$x$ region. This is consistent with an independent analysis based on an effective theory of small-$x$ QCD \cite{Boer:2018vdi,draft}.

\subsection{The result}
Let us introduce the convenient notation 
\beq
F_g\equiv H_g-\frac{\xi^2}{1-\xi^2}E_g, \qquad \tilde{F}_g\equiv \tilde{H}_g-\frac{\xi^2}{1-\xi^2}\tilde{E}_g, \label{conveni}
\eeq
for the linear combinations of GPDs which appear in the Dirac traces (\ref{trac}), (\ref{trac2}) (\ref{trac3}). 
Similarly to (\ref{h2}), we  define ($n=1,2,3$) 
\beq
{\cal F}^{(n)}_g(\xi)=\int_{-1}^1 dx \frac{\xi^{2(n-1)}F_g(x,\xi)}{(x^2-\xi^2)^n},\qquad \tilde{\cal F}^{(n)}_g(\xi)= \int_{-1}^1 dx \frac{\xi^{2(n-2)}x\tilde{F}_g(x,\xi)}{(x^2-\xi^2)^n}, \label{tildemoment}
\eeq
and
\beq
{\cal C}_g^{(2)}(\xi) = \int_{-1}^1dx \frac{\xi^2 x C_g(x,\xi)}{(x^2-\xi^2)^2},
\qquad {\cal C}'^{(2)}_g (\xi)= \int_{-1}^1dx \frac{\xi^2 xC'_g(x,\xi)}{(x^2-\xi^2)^2}. 
\eeq
Using these moments, our final result for the  cross section can be expressed as 
\begin{align}
\dfrac{d\sigma_{\rm DSA}}{dy dQ^{2} d \phi_{l_\perp}dz dq_\perp^2 d^{2}\Delta_\perp }  =\dfrac{e_q^2\alpha_{em}^2\alpha_s^2 y}{2\pi^3Q^2N_c} 
\dfrac{\xi(1+\xi)|l_\perp||\Delta_\perp|\cos(\phi_{l_\perp}-\phi_{\Delta_\perp})}{z\overline{z}(W^{2}+Q^{2})(W^{2}-M_J^{2})(q_\perp^2+\mu^2)^2}{\rm Re}(\Sigma_L+\Sigma_O+\Sigma_h+\Sigma_C), \label{cross}
\end{align} 
where the contribution from the gluon OAM is 
\beq
\Sigma_L &=&  -  \left( {\cal F}^{(1)*}_g     +  4(1-\beta) {\cal F}^{(2)*}_g \right){\cal L}_g^{(2)}+(z-\bar{z})^2{\cal F}_g^{(1)*}\left({\cal L}_g^{(2)}+8(1-\beta){\cal L}_g^{(3)}\right).
\eeq
There are also contributions from the `odderon' 
\beq
\Sigma_O &=&  -  \frac{1}{2}\left( {\cal E}^{(1)*}_g     +  4(1-\beta) {\cal E}^{(2)*}_g \right){\cal O}_g^{(2)}+\frac{(z-\bar{z})^2}{2}{\cal E}_g^{(1)*}\left({\cal O}_g^{(2)}+8(1-\beta){\cal O}_g^{(3)}\right),
\eeq
and from the gluon helicity 
\beq
\Sigma_h &=& (1-\xi){\cal F}_g^{(1)*} \tilde{\cal F}_g^{(2)} 
-(1-\xi)(z-\bar{z})^2\Biggl[ 8(1-\beta)\left({\cal F}_g^{(1)*}+4(1-\beta){\cal F}_g^{(2)*}\right)\tilde{\cal F}^{(3)}_g  \nn  
&& \qquad \qquad  + \left(  (4\beta-3){\cal F}_g^{(1)*}+16(1-\beta){\cal F}^{(2)*}_g+32(1-\beta)^2{\cal F}_g^{(3)*}\right) \tilde{\cal F}^{(2)}_g  \Biggr],
\eeq 
and from the spin-orbit correlation  
\beq 
\Sigma_C = 2(1-\beta) \left(2{\cal C}_g^{(2)*}\tilde{\cal F}_g^{(2)}+\xi {\cal C}'^{(2)*}_g \tilde{\cal E}_g^{(2)} \right).
\eeq
Importantly $\Sigma_h$ contains integrals  with  triple  poles   
\begin{equation}
\begin{split}
& {\cal H}_g^{(3)}(\xi)= \int_{-1}^1 dx \frac{\xi^4 H_g(x,\xi)}{(x^2-\xi^2)^3}, \qquad {\cal E}_g^{(3)}(\xi)= \int_{-1}^1 dx \frac{\xi^4 E_g(x,\xi)}{(x^2-\xi^2)^3}, \label{prob} \\
& \tilde{\cal H}_g^{(3)}(\xi)= \int_{-1}^1 dx \frac{\xi^2x \tilde{H}_g(x,\xi)}{(x^2-\xi^2)^3}, \qquad \tilde{\cal E}_g^{(3)}(\xi)= \int_{-1}^1 dx \frac{\xi^2x\tilde{E}_g(x,\xi)}{(x^2-\xi^2)^3},
\end{split}
\end{equation}
These are similar to the integrals which we encountered before (\ref{lg3}) and are   contained in $\Sigma_L, \Sigma_O$. 
They are well-defined if the second derivative of $H_g$, etc, are continuous at $x=\pm \xi$. Unfortunately, however, this condition is known to be  violated for $H_g$ and $E_g$ at least for asymptotically large values of the renormalization scale $\mu_r$  
\beq
H_g(x,\xi,\mu_r), E_g(x,\xi,\mu_r) \sim \theta(\xi-|x|) (x^2-\xi^2)^2 \qquad (\mu_r\to \infty) \, .
\eeq
 Essentially the same problem was previously encountered in the exclusive production of $P$-wave charmonium states   \cite{Cui:2018jha} where it was concluded that collinear factorization is violated due to this `end-point' singularity. 

In the present problem, clearly the origin of this divergence is the expansion (\ref{expand}) in $k_\perp$ and $\Delta_\perp$. Had we not done this, the cross section would have been  perfectly finite and given in terms of GTMDs (not GPDs). But then we would not be able to exploit the formula (\ref{lgx}) to establish a (direct) connection to gluon OAM. Without the detailed knowledge of the behavior of gluon GPDs at $x=\pm \xi$, it is not clear to us whether the integrals (\ref{prob}) are always divergent for realistic values of $\mu_r$. However, as already noticed in \cite{Bhattacharya:2022vvo}, the integrals with triple poles (\ref{prob}) and (\ref{lg3}) always accompany the prefactor $(z-\bar{z})^2$, hence they all vanish at $z=\frac{1}{2}$. Since by symmetry $z=\frac{1}{2}$ is a stationary point of the dijet cross section, we can expect that the   ${\cal O}(z-\bar{z})^2$ `curvature' corrections around $z=\frac{1}{2}$ is moderate in practice and may be neglected to first approximation once the divergence at $z\neq \frac{1}{2}$ is regularized by the the $k_\perp$-dependent (GTMD) approach.   Further neglecting the odderon contribution and the terms which are explicitly suppressed by powers of $\xi\ll 1$, we arrive at a practically useful formula valid near $z\sim  \frac{1}{2}$ 
\beq
\Sigma_L+\Sigma_O+\Sigma_h+\Sigma_C \approx -  \left( {\cal H}^{(1)*}_g    +4(1-\beta) {\cal H}^{(2)*}_g \right){\cal L}_g^{(2)} +\left({\cal H}_g^{(1)*} + 4(1-\beta) {\cal C}_g^{(2)*}\right) \tilde{\cal H}_g^{(2)}. \label{main}
\eeq
In \cite{Bhattacharya:2022vvo},  we did not calculate the last term related to the spin-orbit correlation assuming that the contribution from the polarized GTMDs and GPDs are suppressed  compared to the unpolarized ones when $\xi\ll 1$.  However, the present complete calculation,  together with the new formula (\ref{g11}),  indicates that this is not the case. From (\ref{lww}) and (\ref{g11}),  we expect that  ${\cal H}_g^{(1)}\sim {\cal H}_g^{(2)}\sim {\cal C}_g^{(2)}$ and $\tilde{\cal H}_g^{(2)}\sim {\cal L}_g^{(2)}$ (see Appendix A). We then immediately see that all the terms in (\ref{main}) are parametrically of the same order.

Let us briefly discuss the impact of the spin-orbit correlation,  deferring a numerical estimate to a later section.  
Previously in  \cite{Bhattacharya:2022vvo},  we have observed  an interesting interplay between the gluon helicity and the gluon OAM in (\ref{main}). As we show in Appendix A,  ${\cal H}_g^{(1,2)}$ are dominated by the imaginary part at small-$\xi$ and there is an approximate relation ${\rm Im}{\cal H}_g^{(1)}(\xi) \approx -2{\rm Im} {\cal H}^{(2)}_g(\xi)$  (cf. Fig.~\ref{cff1}).
Plugging this into (\ref{main}) (without the ${\cal C}_g^{(2)}$ term), we find that the cross section is roughly proportional to 
\beq
d\sigma_{\rm DSA} \sim {\cal H}^{(1)}_g \left(\tilde{\cal H}^{(2)}_g+\frac{q_\perp^2 -z\bar{z}Q^2}{q_\perp^2+z\bar{z}Q^2}{\cal L}_g^{(2)}\right). \label{dsasimp}
\eeq
Since the helicity and OAM have opposite signs at small-$x$ (\ref{lww}), when $q^2_\perp\gg    z\bar{z}Q^2\approx \frac{Q^2}{4}$, the two terms in (\ref{dsasimp}) tend to cancel each other, and when  $q^2_\perp \ll z\bar{z}Q^2$, they  add up. Such an expectation  was  borne out by the numerical analysis performed in \cite{Bhattacharya:2022vvo}. 
However, this scenario is modified by the spin-orbit term. Due to (\ref{cgx}), we expect that ${\cal C}_g^{(2)}\approx -{\cal H}_g^{(2)}$, and this effectively enhances the coefficient of $\tilde{\cal H}_g^{(2)}$ in (\ref{main}) by a factor $1+2(1-\beta)$.  As a result, the impact of the cancellation when $q_\perp^2 \gg Q^2/4$ is diminished unless $\beta$ is close to unity.

\section{Quark exchange channel}
\label{s:quark_channel}
\begin{figure}[htbp]
  \includegraphics[width=0.3\linewidth]{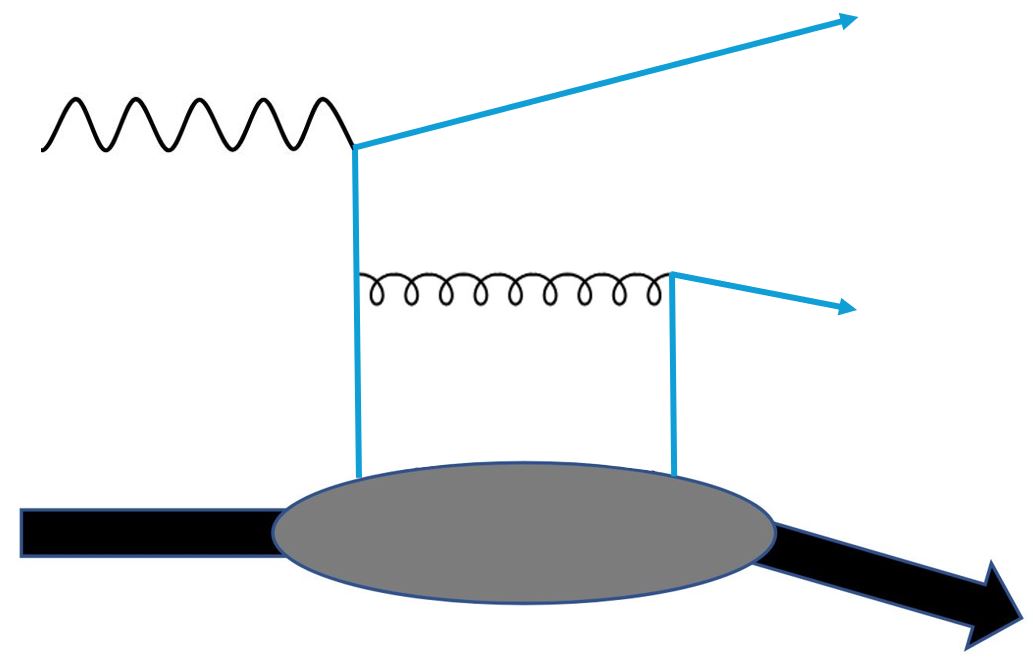} \hspace{2cm}
   \includegraphics[width=0.3\linewidth]{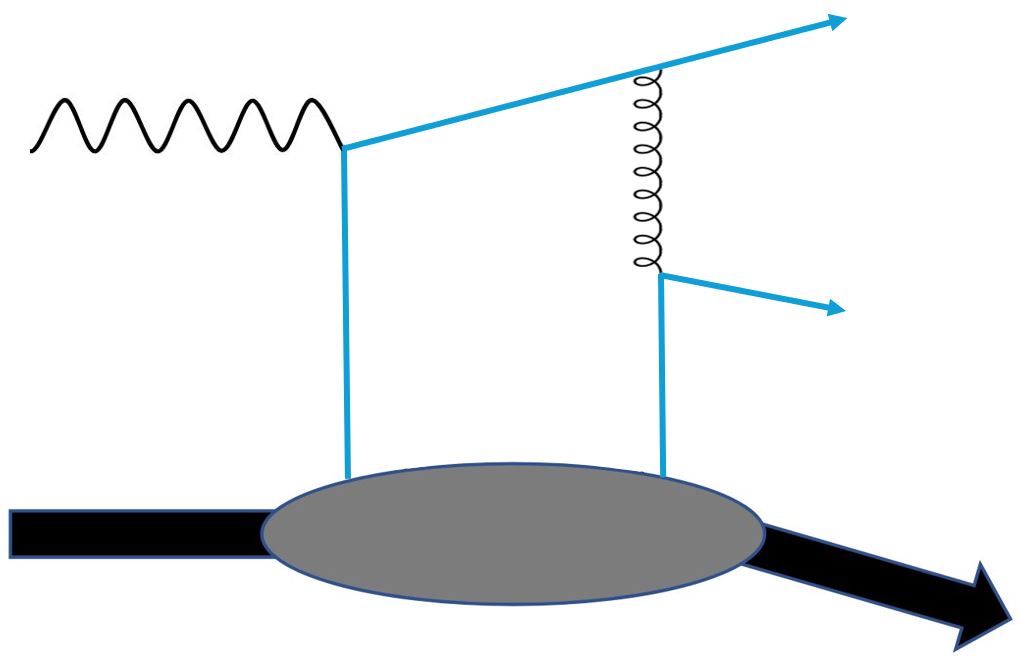}
\caption{Displayed are the leading-order Feynman diagrams illustrating exclusive dijet production initiated by quarks. The corresponding set of two diagrams, with the quark lines exchanged, is implied.}
\label{fig:quarks}
\end{figure} 
\begin{figure}[htbp]
  \includegraphics[width=0.3\linewidth]{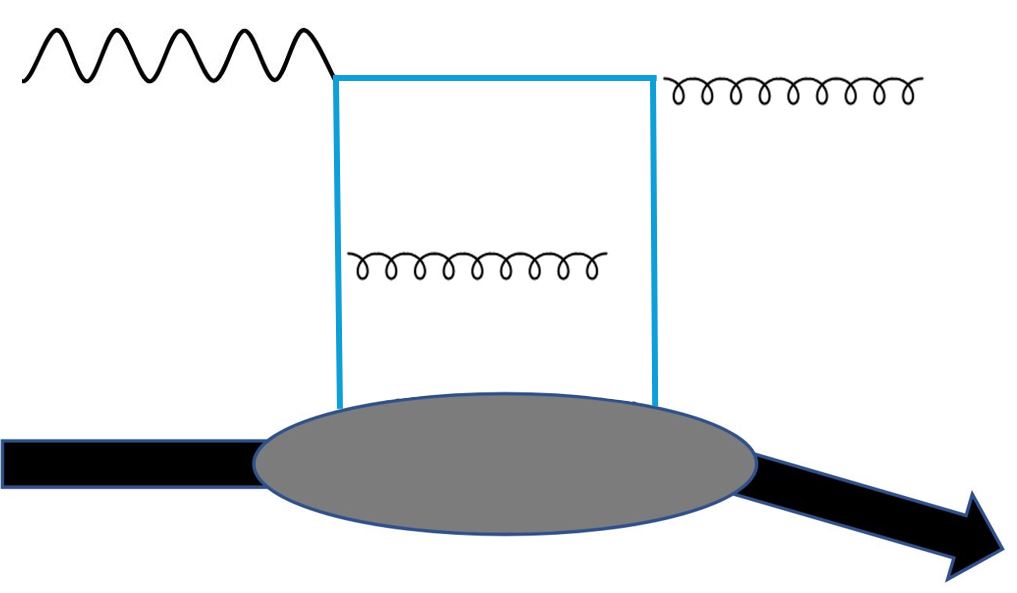} \hspace{2cm}
   \includegraphics[width=0.3\linewidth]{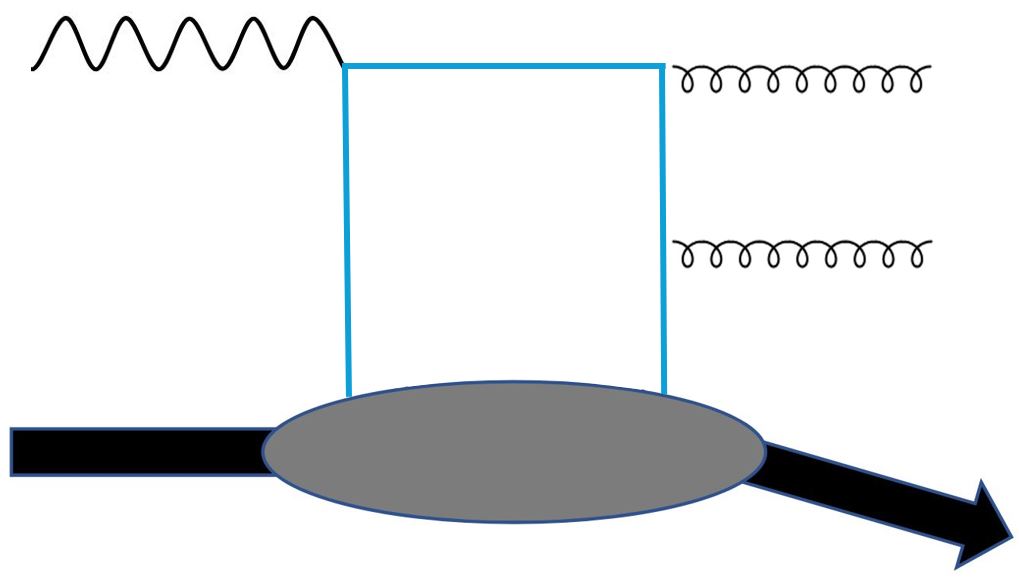}
\caption{Presented here are the leading-order Feynman diagrams showing exclusive gluon-jet production initiated by quarks. The corresponding pair of diagrams, wherein the quark lines are exchanged, is implied.}
\label{fig:quarks2}
\end{figure} 
We now discuss the quark exchange diagrams for exclusive dijet production. Our original goal was to obtain a formula similar to (\ref{cross}) but this time featuring the quark OAM (\ref{quarkgluonoam}). However, we have encountered an unexpected problem whose resolution is beyond the scope of this work.  Therefore, instead of presenting the complete calculation, we will focus on explaining the origin and nature of the problem which may be an interesting observation by itself. 

The relevant diagrams are shown in Fig.~\ref{fig:quarks}. Their sum is proportional to 
\beq
A^q_{L/T}&\propto & \frac{\epsilon_{L/T}^\rho}{(k+\Delta/2+q_2)^2} \bar{u}(q_1)\Biggl[ \gamma^\alpha \frac{\Slash k -\frac{\Slash \Delta}{2} +\Slash q}{(k-\Delta/2+q)^2} \gamma_\rho \langle p'| \psi \bar{\psi}|p\rangle \gamma_\alpha + \gamma_\rho \frac{\Slash q_1-\Slash q}{(q_1-q)^2}\gamma^\alpha \langle p'|\psi\bar{\psi}|p \rangle \gamma_\alpha  \Biggr]v(q_2)  \nn
 &&+ \frac{\epsilon_{L/T}^\rho}{(k-\Delta/2-q_1)^2} \bar{u}(q_1)\Biggl[ \gamma^\alpha\langle p' |\psi\bar{\psi}|p\rangle \gamma_\rho  \frac{\Slash k +\frac{\Slash \Delta}{2} -\Slash q}{(k+\Delta/2-q)^2} \gamma_\alpha  + \gamma^\alpha\langle p'| \psi\bar{\psi}|p\rangle \gamma_\alpha  \frac{-\Slash q_2+\Slash q}{(-q_2+q)^2}\gamma_\rho  \Biggr]v(q_2) .
\eeq 
We use the Fierz identity 
\beq
\langle \psi\bar{\psi}\rangle =  -\frac{1}{4}\langle \bar{\psi}\gamma^+ \psi\rangle \gamma^- +\frac{1}{4}\langle \bar{\psi}\gamma_5\gamma^+ \psi\rangle \gamma_5\gamma^- +\cdots,
\eeq
to project onto the leading twist unpolarized (\ref{gtmd1}) and polarized  (\ref{gtmdqtilde}) quark GTMDs. We then extract the twist-two and twist-three parts by expanding to linear order in $k_\perp$ and $\Delta_\perp$. The twist-two part can be found in \cite{Braun:2005rg}. As for the twist-three part, consider for definiteness  the longitudinally polarized case  $\Slash \epsilon_L \to \epsilon_L^+\gamma^-$.\footnote{ QED gauge invariance is known to be more subtle in quark exchange diagrams, see e.g.,  \cite{Anikin:2000em}. However, we do not expect this issue to resolve the problem that we report here.  } The unpolarized and polarized amplitudes are given by 
\beq
A^{q,{\rm unpol} }_{L}
&\propto & \frac{\langle \bar{\psi}\gamma^+\psi\rangle}{8z\bar{z}\xi (P^+)^2q^-}(\bar{u}\Slash \epsilon_L v) \Biggl[ \frac{z}{x-\xi}+\frac{\bar{z}}{x+\xi}  + \frac{(-z^2+\bar{z}^2)(x^2+\xi^2) -2x\xi (z^2+\bar{z}^2)}{z\bar{z}(x^2-\xi^2)^2} \frac{k_\perp \cdot q_\perp}{P^+q^-} \nn 
&& \qquad \qquad + \frac{(z^2-\bar{z}^2)(x^3-3\xi^2 x) -2\xi^3 (z^2+\bar{z}^2)}{2\xi z\bar{z}(x^2-\xi^2)^2} \frac{q_\perp\cdot \Delta_\perp}{P^+q^-}\Biggr], \label{unq} 
\eeq
\beq
A_{L}^{q,{\rm pol}} 
&\propto& -\frac{\langle \bar{\psi}\gamma_5\gamma^+\psi\rangle}{8z\bar{z}\xi (P^+)^2q^-}(\bar{u}\Slash \epsilon_L \gamma_5 v) 
 \Biggl[ \frac{z}{x-\xi}+\frac{\bar{z}}{x+\xi}  + \frac{(-z^2+\bar{z}^2)(x^2+\xi^2) -2x\xi (z^2+\bar{z}^2)}{z\bar{z}(x^2-\xi^2)^2} \frac{k_\perp \cdot q_\perp}{P^+q^-} \nn 
&& \qquad \qquad  + \frac{(z^2-\bar{z}^2)(x^3-3\xi^2 x) -2\xi^3 (z^2+\bar{z}^2)}{2\xi z\bar{z}(x^2-\xi^2)^2} \frac{q_\perp\cdot \Delta_\perp}{P^+q^-}\Biggr]. \label{polq}
\eeq
 Integrating the ${\cal O}(k_\perp)$ term of (\ref{unq}) and (\ref{polq}) over $k_\perp$, we obtain the GPD $L_g(x,\xi)$ associated with the quark OAM (\ref{lgx0}) and the quarks' spin-orbit correlation  \cite{Lorce:2011kd}, cf., (\ref{orbit})
 \beq
C_q(x,\xi)\equiv  \int d^2k_\perp \frac{k_\perp^2}{M^2}G_{1,1}^q(x,\xi). 
\eeq 
On the other hand, from the ${\cal O}(\Delta_\perp)$ term of (\ref{unq}) and (\ref{polq}), we obtain the unpolarized and polarized quark GPDs $H_q$, $\tilde{H}_q$. Up to this point, the discussion is completely parallel with the gluon exchange channel in the previous section. The difference starts to appear after the subsequent integration over $x$ which  gives two types of moments   with double poles at $x=\pm \xi$  
\beq
&& (z^2-\bar{z}^2) \int_{-1}^1 dx \frac{  L_q(x,\xi)}{(x^2-\xi^2)^2}= \frac{z^2-\bar{z}^2}{2} \int_{-1}^1 dx \frac{  L^{(+)}_q(x,\xi)}{(x^2-\xi^2)^2}, \label{int1} \\
&& (z^2+\bar{z}^2) \int_{-1}^1 dx \frac{x L_q(x,\xi)}{(x^2-\xi^2)^2} = \frac{z^2+\bar{z}^2}{2} \int_{-1}^1 dx \frac{ x L^{(-)}_q(x,\xi)}{(x^2-\xi^2)^2}, \label{br}
\eeq
\beq
&& (z^2-\bar{z}^2)\int_{-1}^1 dx \frac{xH_q(x,\xi)}{(x^2-\xi^2)^2} = \frac{z^2-\bar{z}^2}{2} \int_{-1}^1 dx \frac{xH^{(+)}_q(x,\xi)}{(x^2-\xi^2)^2}, \label{vector} \\
&&(z^2+\bar{z}^2)\int_{-1}^1 dx \frac{H_q(x,\xi)}{(x^2-\xi^2)^2} =\frac{z^2+\bar{z}^2}{2} \int_{-1}^1 dx \frac{H^{(-)}_q(x,\xi)}{(x^2-\xi^2)^2}, 
 \label{br2}
\eeq
where we defined the $C$-parity even $(+)$ and odd $(-)$ quark GPD combinations as 
\beq
&& L_q^{(\pm )}(x,\xi) = L_q(x,\xi)\pm L_q(-x,\xi), \nn
&& H^{(\pm)}_q(x,\xi)=H_q(x,\xi)\mp H_q(-x,\xi).
\eeq 
The $C$-odd contributions arise because the nucleon is not an eigenstate of $C$-parity and the outgoing $q\bar{q}$ pair can be in both $C$-even and $C$-odd states. 
$L_q^{(+)}$ is directly related to the nucleon spin sum rule, while $L_q^{(-)}$ can be understood as the OAM of valence quarks.  

The problem is that, in the quark channel, it is the double pole that is dangerous for factorization. Indeed, the above integrals are finite provided the first derivative of $H_q^{(\pm)},L^{(\pm)}_q$ is continuous at $x=\pm \xi$. This condition is known to be violated for $H_q^{(\pm)}$ (see \cite{Diehl:2003ny} and references therein), and presumably  also for $L_q^{(\pm)}$. Note that (\ref{vector}) is exactly the same integral that  violates the collinear factorization of the exclusive production of a transversely polarized vector meson \cite{Mankiewicz:1999tt,Anikin:2002wg}. This integral, together with (\ref{int1}),  can be eliminated by setting $z=\frac{1}{2}$ as in the gluon case. Surprisingly, however, the $C$-odd integrals (\ref{br}), (\ref{br2})  survive  even after this approximation and contribute to the cosine modulation (\ref{ang}) after interfering with the twist-two transverse amplitude $A_T^q$ \cite{Braun:2005rg}.  We have seen in the previous section that in the gluon case the $C$-odd odderon exchange does not break factorization at least around $z\approx \frac{1}{2}$. Thus the problem is unique to the quark exchange (in contrast to \cite{Nabeebaccus:2023rzr}) and exemplifies a novel pattern of factorization breaking. 

Another class of diagrams with quark exchange in the $t$ channel could also contribute: diagrams where both jets are initiated by hard gluons (Fig.~\ref{fig:quarks2}). In that case, by color conservation the gluon pair is on a singlet state and it is thus $C$-even. As a result, the $C$-odd part of the quark distribution is probed and we face the same non-cancellation of double poles even for $z\approx\frac{1}{2}$.

It is expected that,  at high energy where $\xi\ll 1$, the quark contribution is subleading compared to the gluon one calculated in the previous section. We therefore neglect it in the numerical analysis below.   However, the problem just described appears to pose a challenge if one would like to probe the quark OAM in the large-$x$ region. 

\section{Semi-inclusive diffractive cross section}
\label{s:siddis}
In a previous publication \cite{Bhattacharya:2022vvo}, we have presented a numerical evaluation of the formula (\ref{main}) (without the last term). While the result cleanly demonstrated the salient features of the cross section such as the  interplay between the gluon OAM and helicity contributions, a serious practical challenge was the overall smallness of the cross section. Considering the luminosity of the Electron-Ion Collider, one has to probe the region of jet transverse momentum around  $q_\perp=1\sim 2$ GeV. The problem is that it is quite difficult to reconstruct jets at such low transverse momentum.  One possibility to circumvent this difficulty is to consider dihadron or heavy-quark pair production  instead of light-quark dijet production (see, e.g., \cite{ReinkePelicer:2018gyh,Bergabo:2022tcu,Linek:2023kga,Fucilla:2022wcg}). 
In this section, we adopt a different strategy  inspired by a recent work  \cite{Hatta:2022lzj}. 

One of the  advantages  of our proposal (as opposed to the earlier proposals   \cite{Ji:2016jgn,Hatta:2016aoc}) is  that the angular modulation $\cos (\phi_{l_\perp}-\phi_{\Delta_\perp})$ we look for  does not depend on the jet azimuthal angle $\phi_{q_\perp}$. In fact, $\phi_{q_\perp}$ has been integrated over to arrive at the formula (\ref{cross}). It is then tempting to integrate also over $q_\perp$. Let us explore this possibility.  
Experimentally, the parameter $\beta$ introduced in (\ref{beta}),  or the invariant mass of the diffractively produced system 
\beq
M_X^2=\frac{q_\perp^2}{z\bar{z}} = \frac{1-\beta}{\beta}Q^2, \label{mx}
\eeq
is easy to measure since it is not strongly affected by the hadronization effects and does not require jet reconstruction. 
Moreover,  at fixed energy and $Q^2$, $\beta$ is in one-to-one correspondence with the skewness parameter 
\beq
\xi=\frac{Q^2}{(2\beta-1)Q^2+2\beta W^2},
\eeq
upon which the Compton amplitudes, ${\cal H}_g^{(1)}(\xi)$ etc., depend. The simplest observable would then be the inclusive cross section $d\sigma/d\beta$ in which  $\beta$ is fixed by inserting the constraint 
\beq
1=\int_0^1 d\beta \delta\left(\beta - \frac{z(1-z)Q^2}{q_\perp^2+z(1-z)Q^2}\right), \label{dec1}
\eeq
and  $q_\perp$ and  $z$ are integrated over  
\beq
\frac{d\sigma}{dy dQ^2 d\phi_{l_\perp}d^2\Delta_\perp d\beta} &=& \int_0^1  dz \int_0^\infty dq_\perp^2 \delta\left(\beta - \frac{z(1-z)Q^2}{q_\perp^2+z(1-z)Q^2}\right)f(z,q_\perp) \nn
&=& \left.\int_0^1 dz \frac{z(1-z)Q^2}{\beta} f(z,q_\perp)\right|_{q_\perp^2=z(1-z)\frac{1-\beta}{\beta}Q^2},
\eeq
where $f$ is the right-hand-side of (\ref{cross}). This is similar to the inclusive diffractive DIS cross section. However, the integral probes the so-called `aligned jet' region $z,\bar{z}\to 0$ where our approximation $z\approx \frac{1}{2}$ breaks down. Besides, at fixed $\beta$, $z,\bar{z}\to 0$ means $q^2_\perp\to 0$ and $\mu^2=z\bar{z}Q^2\to 0$, so the twist expansion such as (\ref{expand}) becomes invalid. Furthermore, already the unpolarized cross section integrated over $z,q_\perp$ receives a significant contribution from this nonperturbative region, hence the calculation is not reliable \cite{Bartels:1996ne,Forshaw:1997dc}.

To remedy this, we need to impose a lower cutoff in $q_\perp $. Since we try to avoid having to measure jets, in practice, we shall impose this cutoff for the transverse momentum $P_{h\perp}$ of certain hadron species $h$ by introducing the corresponding fragmentation function. We are thus led to consider  the cross section
\beq
\frac{d\sigma}{dy dQ^2 d\phi_{l_\perp}dz_fdP_{h\perp}^2 d^2\Delta_\perp} =\int_{z_f}^1 \frac{dz}{z} \frac{z^2}{z_f^2}D_q^{h}\left(\frac{z_f}{z}\right) f\left(z,\frac{z}{z_f}P_h\right)+\int_{z_f}^1 \frac{d\bar{z}}{ \bar{z}}\frac{\bar{z}^2}{z_f^2}D_{\bar{q}}^{h}\left(\frac{z_f}{\bar{z}}\right) f\left(z,\frac{\bar{z}}{z_f}P_h\right) ,
\eeq
where 
\beq
z_f=\frac{P\cdot P_h}{P\cdot q} ,
\eeq 
is the standard variable in semi-inclusive DIS. 
 We then insert (\ref{dec1}) and integrate over $P_h^2$ with a lower cutoff $\Lambda\sim 1$ GeV. 
\beq
\frac{d\sigma}{dy dQ^2 d\phi_{l_\perp}dz_f d\beta d^2\Delta_\perp} &=&  \int_{\Lambda^2}^\infty dP_h^2\int_{z_f}^1 \frac{dz}{ z} \frac{z^2}{z_f^2}D_q^{h}\left(\frac{z_f}{z}\right) f\left(z,\frac{z}{z_f}P_h\right)\delta\left(\beta - \frac{z(1-z)Q^2}{q_\perp^2+z(1-z)Q^2}\right)+(q\to \bar{q}) \nn
&=& \int^1_{z_f} \frac{dz}{z}\int^\infty _{\Lambda^2\frac{z^2}{z_f^2}} dq_\perp^2 D_q^{h}\left(\frac{z_f}{z}\right)  f\left(z,q_\perp\right)\delta\left(\beta - \frac{z(1-z)Q^2}{q_\perp^2+z(1-z)Q^2}\right) +(q\to \bar{q})  \nn
&=&
\left.\int^{z_{max}}_{z_f} \frac{dz}{ z}\frac{z(1-z)Q^2}{\beta^2}\left(D_q^{h}\left(\frac{z_f}{z}\right)+D_{\bar{q}}^h\left(\frac{z_f}{z}\right)\right)  f(z,q_\perp)\right|_{q_\perp^2=z(1-z)\frac{1-\beta}{\beta}Q^2}, \label{siddis} 
\eeq
where the upper limit of the $z$-integral is fixed by the condition  
\beq
q_\perp^2=z(1-z)\frac{1-\beta}{\beta}Q^2 > \Lambda^2 \frac{z^2}{z_f^2} \qquad \to \qquad z_f<z< \frac{ \frac{1-\beta}{\beta}Q^2}{\frac{1-\beta}{\beta}Q^2+\frac{\Lambda^2}{z_f^2}} \equiv z_{max} \, .
\eeq
Requiring  $z_{max}>z_f$, we find the following  constraints on the parameters involved
\beq
\frac{\beta \Lambda^2}{(1-\beta)Q^2} < z_f(1-z_f)<\frac{1}{4}.
\eeq
Another constraint specific to the present work is that 
we have to make sure that the $z$-integral is limited to the region $z\sim \frac{1}{2}$. This can be easily done. For example, taking  $Q^2=3$ GeV$^2$, $\beta=0.2$, $\Lambda^2=1$ GeV$^2$, $z_f=0.4$, we find 
\beq
0.4<z<0.66. 
\eeq

The formula (\ref{siddis}) is an example of semi-inclusive diffractive DIS, or `SIDDIS' introduced recently in \cite{Hatta:2022lzj} (see also \cite{Fucilla:2023mkl,Guo:2023uis}). It is a combination of semi-inclusive DIS (SIDIS) and diffractive DIS. One measures an elastically scattered nucleon in the final state followed by a rapidity gap $\Delta Y= \ln \frac{1}{2\xi}$ devoid of particles, and then by a diffractively produced system with invariant mass $M_X$  which spans a rapdity interval $\Delta Y=\ln \frac{1}{\beta}$. Out of this diffractive system, one inclusively  measures the transverse momentum of one hadron species (above some cutoff $P_{h\perp}>\Lambda$ in our case).  For the present purpose, one may consider tagging  several hadron species, or even all charged hadrons, in order to increase the cross section. Experimentally, this is a crucial simplification  because measuring hadrons is  easier than measuring jets at low transverse momentum.

\section{Numerical results}
\label{s:numerics}
Finally, in this section we make predictions for the EIC. The numerical result for the dijet cross section has been reported in \cite{Bhattacharya:2022vvo}. Here we explain the details of the computation and study the impact of the spin-orbit contribution which was not included in  \cite{Bhattacharya:2022vvo}. We also make new predictions for the `SIDDIS' observable discussed in the previous section. 

\subsection{Non-perturbative input: Double distribution approach}
In order to evaluate the various Compton form factors in the formula (\ref{main}), first we have to model the GPDs $f(x,\xi)=\{H_g(x,\xi),\tilde{H}_g(x,\xi),xL_g(x,\xi),xC_g(x,\xi)\}$. We use the so-called double distribution approach \cite{Radyushkin:1998es,Radyushkin:2000uy}  wherein a GPD $f(x,\xi)$ is reconstructed from its forward limit $f(x)$ (PDF) via the formula
\begin{align}
f(x,\xi) & = \dfrac{1}{\xi} \theta(x-\xi) \int^{x_1}_{x_2} d\beta K(\dfrac{x-\beta}{\xi},\beta) + \dfrac{1}{\xi}\theta(\xi-x) \left[ \int^{x_1}_0 d\beta K(\dfrac{x-\beta}{\xi}, \beta) - (x\to -x)\right],
\end{align}
where
\begin{equation}
K(\alpha,\beta) = \frac{15}{16} \frac{((1-|\beta|)^2-\alpha^2)^2}{(1-|\beta|)^5} \times f(\beta), 
\end{equation}
and the limits of the integrals $(x_1, x_2, x_3)$ are:
\begin{align}
 x_1=\dfrac{x+\xi}{1+\xi}, \quad x_2 = \dfrac{x-\xi}{1-\xi}, \quad x_3=-\dfrac{\xi-x}{1+\xi} .
\end{align}
Once the GPDs are obtained this way, the associated Compton form factors are evaluated using the formulas collected  in Appendix A. 
We use $H_g(x)=x G(x)$ from JAM 19 (NLO) \cite{Sato:2019yez} and $\tilde{H}_g(x)=x \Delta G(x)$ from JAM 17 (NLO)~\cite{Ethier:2017zbq}. As for $xL_g(x)$ and $xC_g(x)$, we use the Wandzura-Wilczek approximation (see \cite{Hatta:2012cs} and (\ref{g11}))
\begin{equation}
\begin{split}
&L_g(x) \, \, \approx \, \, x \int^{1}_x \dfrac{dx'}{x'^2} (H_g (x') + E_g (x')) \, - \, 2 x \int^{1}_x \dfrac{dx'}{x'^2} \Delta G (x'), \\[0.2cm]
& C_g(x) \, \, \approx \, \,  x\int_x^{1}\frac{dx'}{x'^2}\tilde{H}_g(x') -2x \int_x^{1} \frac{dx'}{x'^2}G(x') ,
\end{split}
\label{lc}
\end{equation}
though of course, we are eventually interested in constraining the total OAM distribution including the genuine twist-three terms.   
In \cite{Bhattacharya:2022vvo}, we neglected $E_g(x)$ in the above formula. However,  a recent study has revealed a rapid growth of $E_g(x)$ at small-$x$, with the ratio of $E_g(x)/H_g(x)$ approaching a constant as $x\to 0$ \cite{Hatta:2022bxn}. We thus include it assuming the form $E_g(x) \approx -0.15H_g(x)$ where the coefficient is from the light-cone spectator model  \cite{Tan:2023kbl}. We however find that the impact of $E_g(x)$ is minor, since $L_g(x)$ in (\ref{lc}) is dominated by the last term as mentioned already. 

We integrate  the cross section over  $|\Delta_\perp|$ assuming a Gaussian form  $d\sigma \propto e^{-b\Delta_\perp^2}$ with $b=5$ GeV$^{-2}$  \cite{Braun:2005rg} and plot 
\beq
\frac{d\sigma_{\rm DSA}}{dy dQ^2 d\Delta \phi dz (d\xi \ {\rm or }\  q_\perp^2)}, 
\qquad  \frac{d\sigma_{\rm DSA}}{dy dQ^2 d\Delta \phi dz_f d\beta},
\label{e:plot}
\eeq
for the dijet and SIDDIS cross sections, respectively. The conversion $\xi \leftrightarrow q_\perp^2$ is done according to (\ref{xiq}). We fix $y=0.7$, $\sqrt{s}=120$ $\rm{GeV}$, $\Delta\phi=\phi_{l_\perp}-\phi_{\Delta_\perp}=0$, and employ three values of photon virtuality $Q^2=2.7$ GeV$^2$, 4.8 GeV$^2$ and  10 GeV$^2$.  Typically, with these values we probe $\xi \big |_{\rm min.} \sim 10^{-3}$ corresponding to  $q_\perp \big |_{\rm min.}=1$ $\rm{GeV}$. 

\subsection{Results for dijet process}
\begin{figure}[htbp]
  \includegraphics[width=0.3\linewidth]{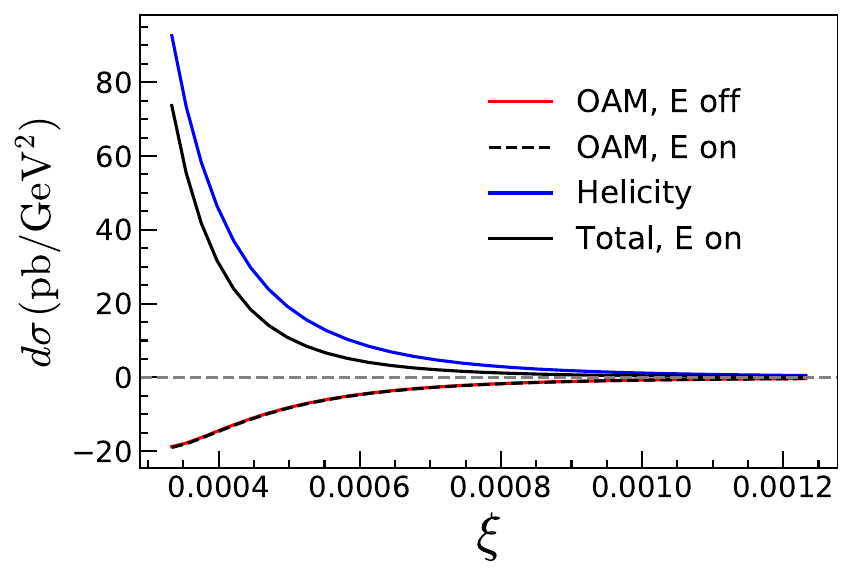}
   \includegraphics[width=0.3\linewidth]{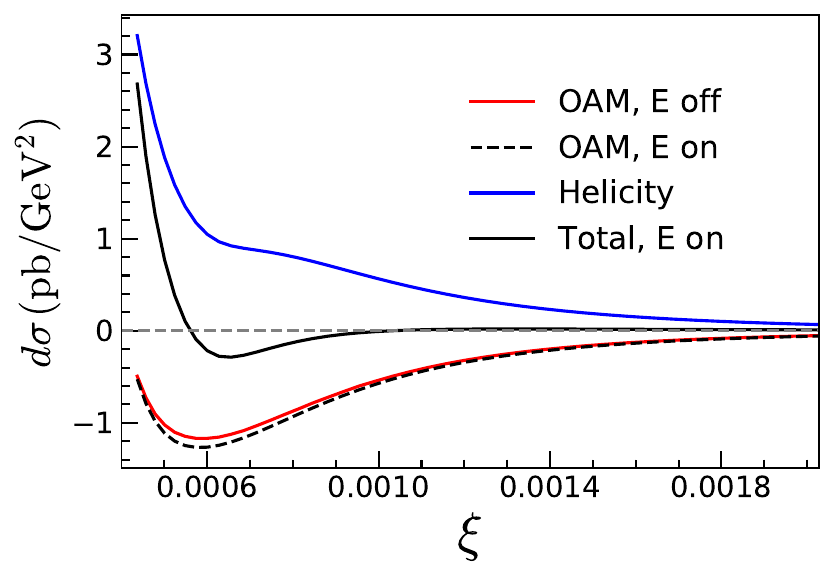}
   \includegraphics[width=0.3\linewidth]{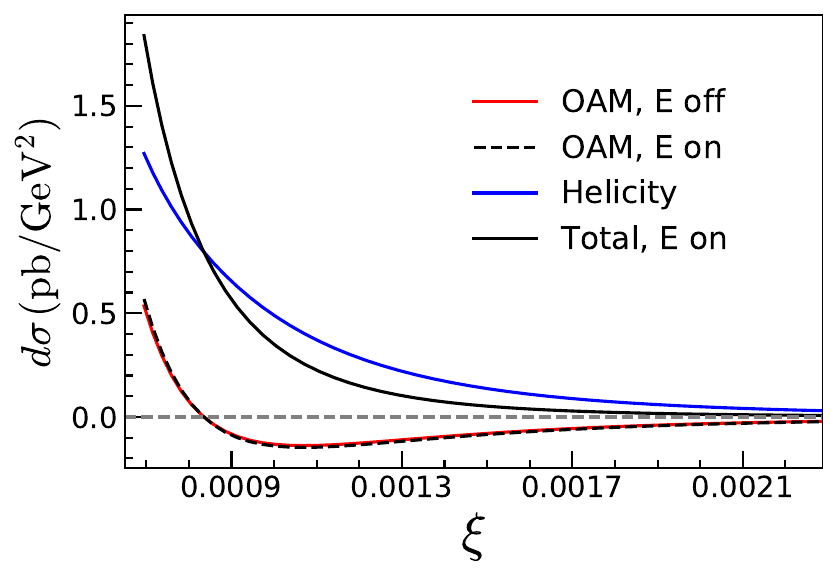}
\caption{The DSA part of the differential cross-section, both with and without $E_g$, is presented as a function of $\xi$ at $Q^2=2.7 \, \textrm{GeV}^2$ (left), $Q^2=4.8 \, \textrm{GeV}^2$ (middle), and $Q^2=10 \, \textrm{GeV}^2$ (right). The results with $E_g =0$ correspond to those reported in our previous work \cite{Bhattacharya:2022vvo}. These plots suggest that the incorporation of contributions from $E_g$ into the orbital angular momentum does not result in a significant alteration of our previous results.}
\label{fig1}
\end{figure} 
\begin{figure}[htbp]
  \includegraphics[width=0.3\linewidth]{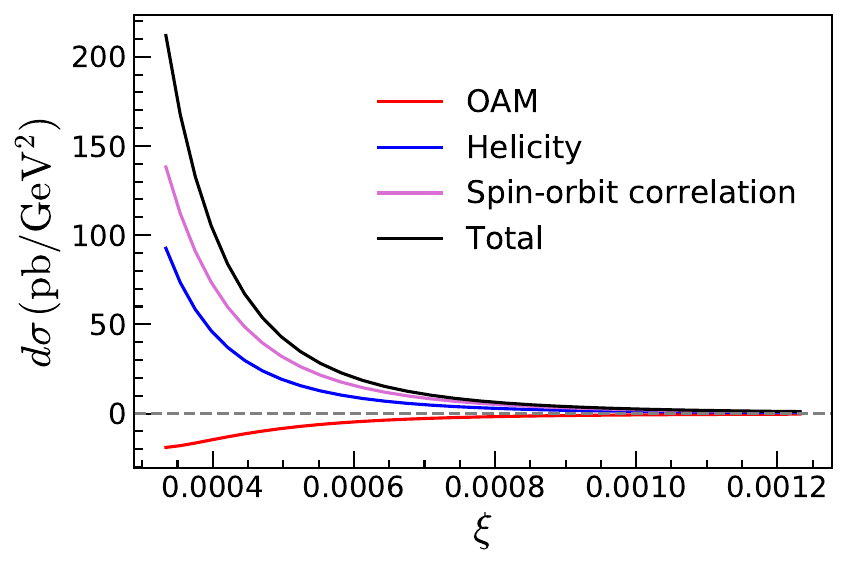}
   \includegraphics[width=0.3\linewidth]{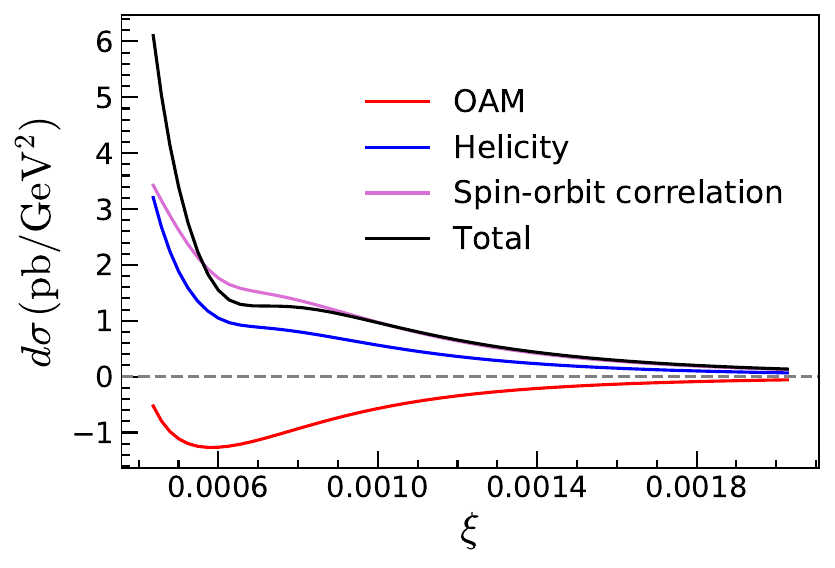}
   \includegraphics[width=0.3\linewidth]{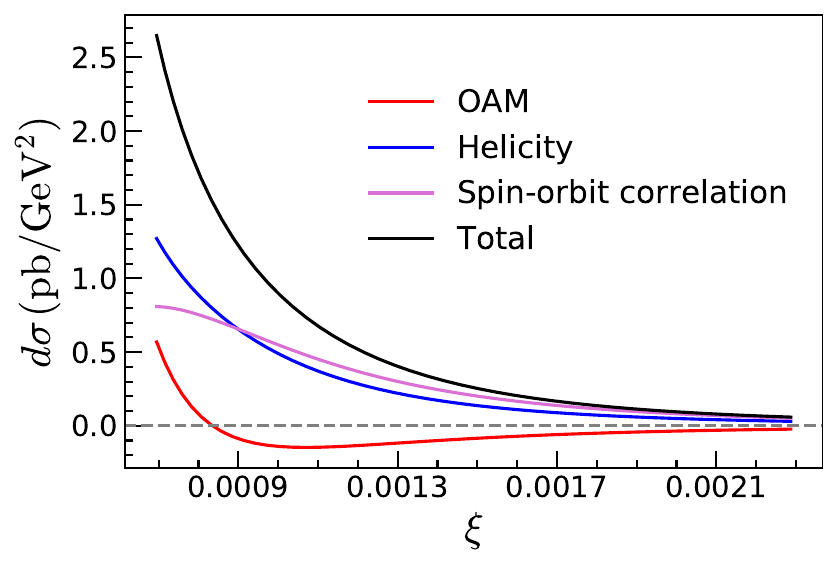}
\caption{The DSA part of the differential cross-section with $E_g \neq 0$ is presented as a function of $\xi$ at $Q^2=2.7 \, \textrm{GeV}^2$ (left), $Q^2=4.8 \, \textrm{GeV}^2$ (middle), and $Q^2=10 \, \textrm{GeV}^2$ (right). These plots include contributions from spin-orbit correlation, explicitly demonstrating how the results in the above figure are quantitatively modified in the presence of this contribution.}
\label{fig2}
\end{figure} 
\begin{figure}[htbp]
  \includegraphics[width=0.3\linewidth]{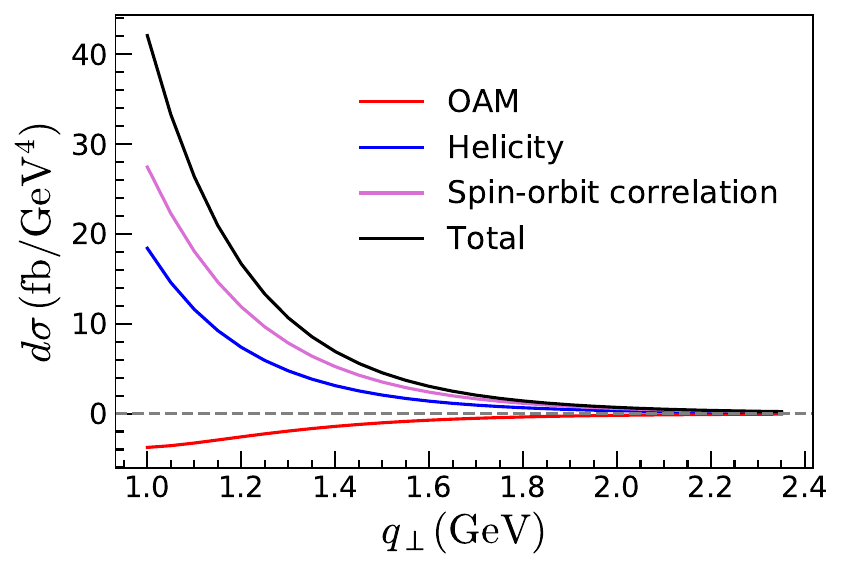}
  \includegraphics[width=0.3\linewidth]{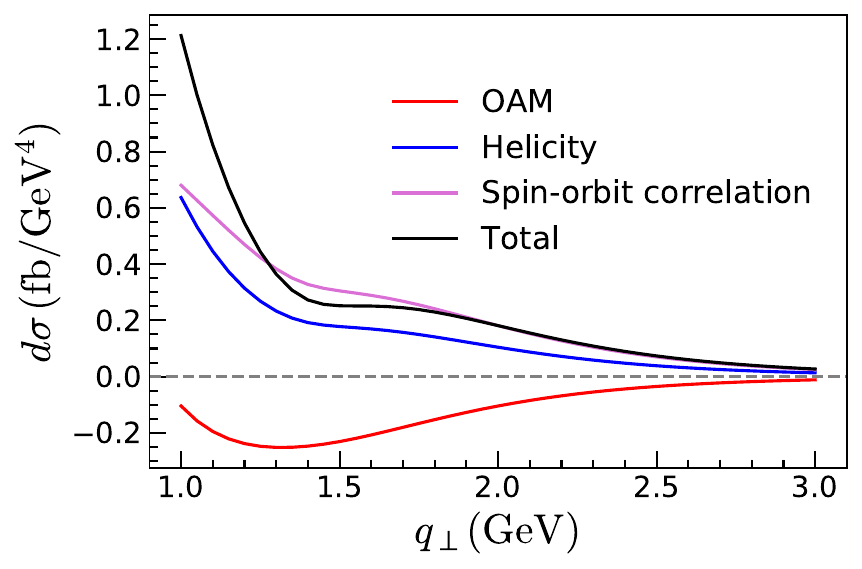}
  \includegraphics[width=0.3\linewidth]{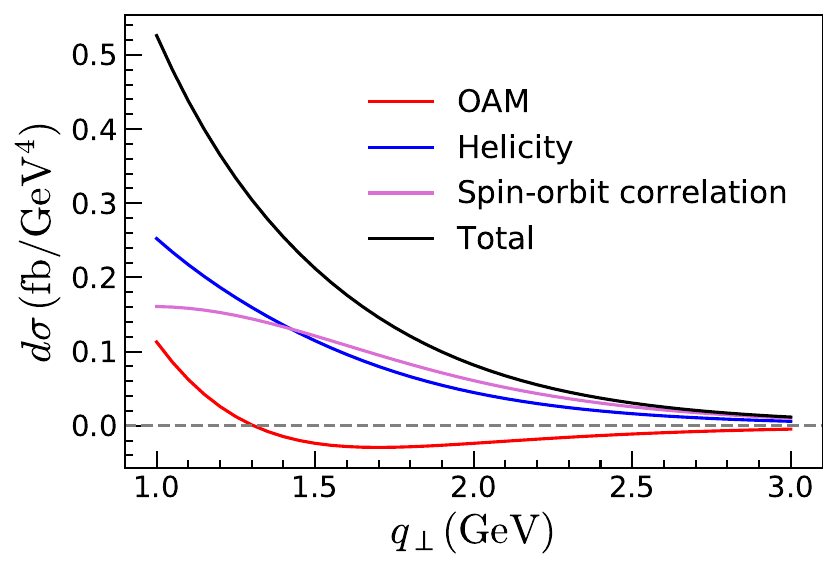}
\caption{The DSA part of the differential cross-section with $E_g \neq 0$ is presented as a function of $q_\perp$ at $Q^2=2.7 \, \textrm{GeV}^2$ (left), $Q^2=4.8 \, \textrm{GeV}^2$ (middle), and $Q^2=10 \, \textrm{GeV}^2$ (right).}
\label{fig3}
\end{figure} 
In Fig.~\ref{fig1} and Fig.~\ref{fig2}, we present Eq.~(\ref{e:plot}) as functions of $\xi$.  In the former, we display the results of our previous work with $E_g$ activated, contrasting them with our published findings. In the latter, we introduce the spin-orbit correlation term. Our analysis reveals that the OAM and helicity contributions tend to either cancel or reinforce each other, depending on $Q^2$, as discussed earlier. Notably, the spin-orbit correlation and helicity terms exhibit comparable magnitudes and are substantial. Since the spin-orbit correlation is solely dictated by the unpolarized gluon distribution, it is relatively better constrained than OAM, which requires precise knowledge of both unpolarized and polarized gluon distributions. The latter has significant uncertainties, particularly at small $x$. Subtracting the contributions from the helicity and spin-orbit correlation terms could provide direct sensitivity to OAM.

In Fig.~\ref{fig3}, we present the same plot as in Fig.~\ref{fig2}, but now as a function of the dijet transverse momentum $q_\perp$. Note that the units of the vertical axis have changed pb $\to$ fb (apart from the trivial dimensional factor GeV$^{-2}$). This is because of the typical value $\xi \sim 10^{-3}$ probed in this kinematics.

It is important to emphasize that our numerical estimations incorporated several approximations, including systematic errors inherent in the double distribution method and the Wandzura-Wilczek approximation. However, the primary source of uncertainty lies in the unpolarized and especially  polarized gluon distributions which are currently not well constrained at small-$x$. Moreover, ideally the next-to-leading order corrections should be included \cite{Boussarie:2016ogo}. Due to these reasons, we did not attempt to include error bars in the above plots.
Therefore, our numerical results provide only a rough estimate of the overall magnitudes and signs of various contributions. In the future, as more accurate determinations of $\Delta G(x)$ become available, our predictions can be correspondingly improved.

\subsection{Results for SIDDIS process}
\begin{figure}[htbp]
  \includegraphics[width=0.55\linewidth]{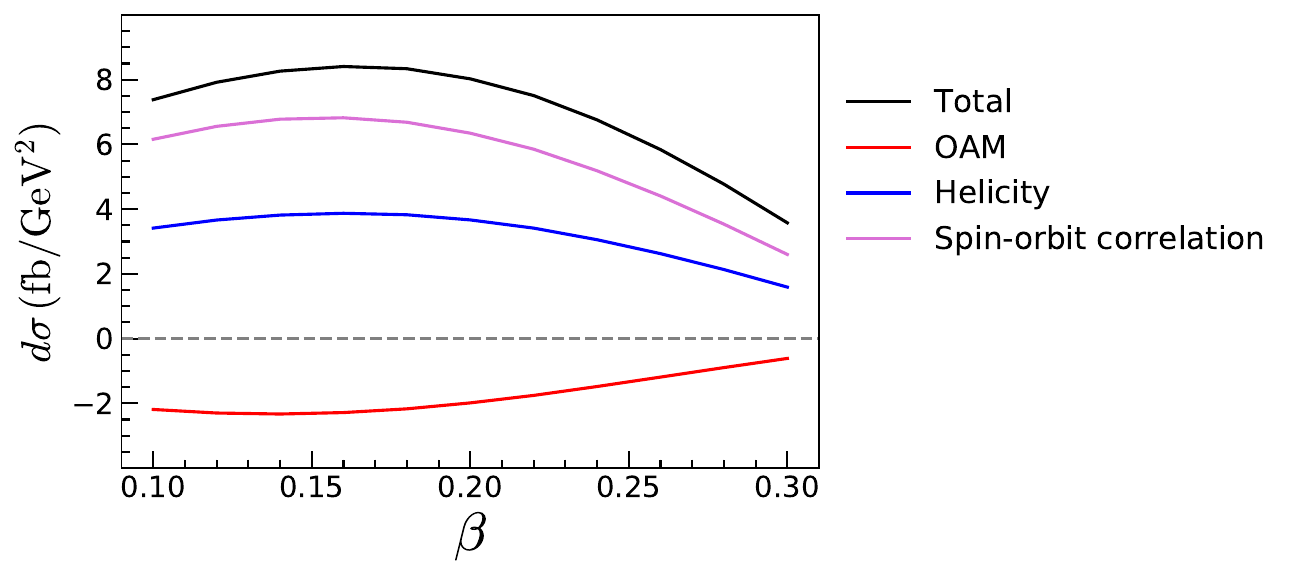}
   \includegraphics[width=0.37\linewidth]{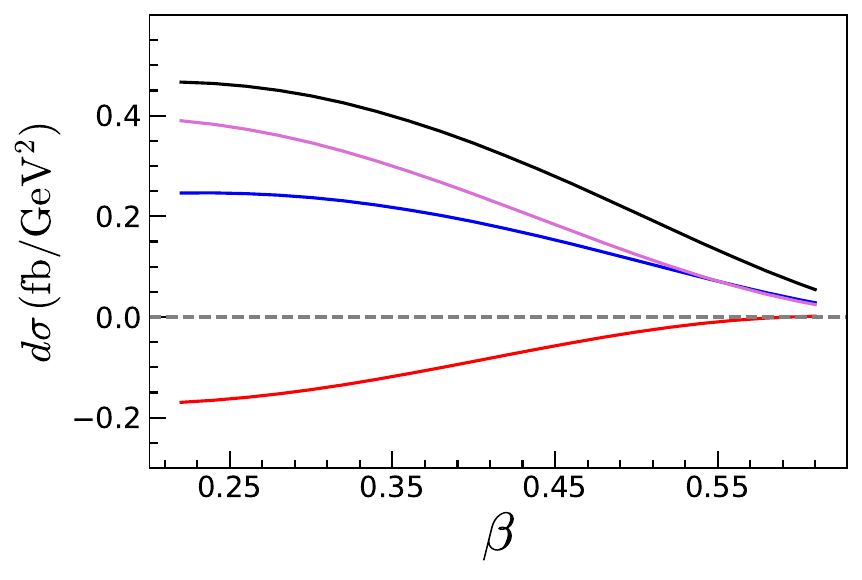}
\caption{Plot depicting the SIDDIS cross section as a function of the $\beta$ parameter for $\Lambda=1$ GeV and at two different $Q^2$ values: $2.7 \, \textrm{GeV}^2$ (left) and $10 \, \textrm{GeV}^2$ (right). The maximum $\beta$ value is restricted by the condition $q_\perp < 1$ GeV.}
\label{fig4}
\end{figure} 
\begin{figure}[htbp]
  \includegraphics[width=0.55\linewidth]{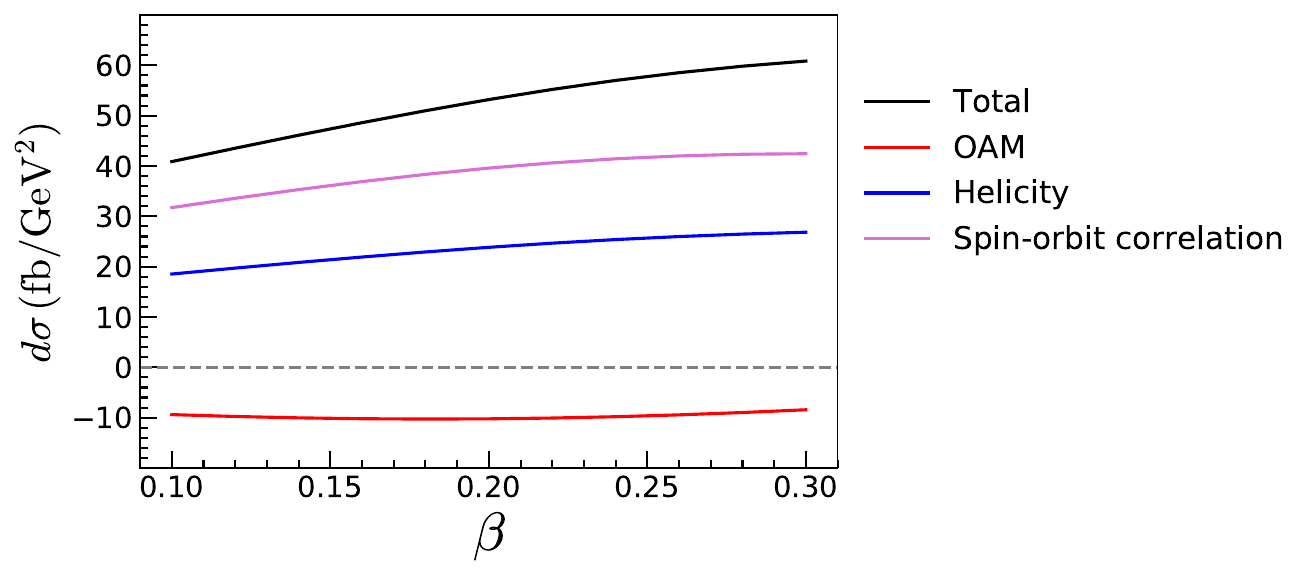}
   \includegraphics[width=0.37\linewidth]{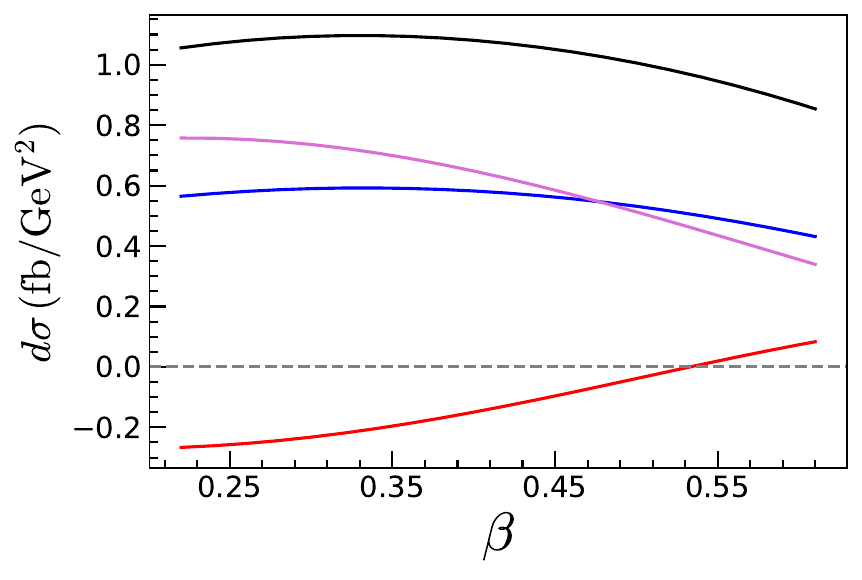}
\caption{Plot depicting the SIDDIS cross section as a function of the $\beta$ parameter for $\Lambda=0.5$ GeV and at two different $Q^2$ values: $2.7 \, \textrm{GeV}^2$ (left) and $10 \, \textrm{GeV}^2$ (right). The change of OAM sign, resulting from the interplay between OAM and helicity at small $x$, is once again clearly observable in the high $\beta$ region.}
\label{fig5}
\end{figure} 
In Figs.~\ref{fig4} and \ref{fig5}, we present the SIDDIS cross section as a function of the $\beta$ parameter for two distinct cutoff values: $\Lambda=1.0$ and $0.5$ GeV. The calculation considers tagged hadron species, such as charged pions and kaons, while the fragmentation quarks are assumed to be $u, \bar{u}, d, \bar{d}, s$ and $\bar{s}$. The change in the sign mechanism of the OAM, as observed in the dijet process, is also evident here for $\Lambda=0.5$ GeV and in the higher regions of $\beta$. While the obtained number values suggest that measuring this process might still be challenging, our main point is that there is no need to reconstruct the dijet transverse momentum, yet sensitivity to OAM is retained. Therefore, this approach may offer a promising route for measurement at the EIC to constrain OAM.

\section{Conclusion}
\label{s:conclusion} 
In conclusion, our overarching goal has been to identify and establish  `practical' physical observables capable of directly measuring parton orbital angular momentum (OAM).  
In this endeavor, we have expanded upon our previous work \cite{Bhattacharya:2022vvo} by offering a comprehensive analysis of the longitudinal double spin asymmetry in exclusive dijet production during electron-proton scattering. Our investigation reaffirms the sensitivity of the $\cos \phi$ angular correlation as a crucial indicator of gluon OAM at small-$x$, and its intricate interplay with gluon helicity. Furthermore, our study uncovers an additional aspect  unexplored in our prior research: the spin-orbit correlation of gluons $C_g(x)$. For the first time, we shed light on the small-$x$ behavior of this distribution (see also \cite{draft}) and find that it is approximately proportional to the unpolarized gluon distribution but with an opposite sign. Consequently, its contribution to the present observable is unsuppressed, contrary to naive expectations.  
The determination of OAM therefore necessitates accurate knowledge of the unpolarized gluon distribution at small-$x$, in addition to the polarized gluon distribution $\Delta G(x)$. Alternatively, since the spin-orbit correlation exists   already in unpolarized hadrons, one may be able to constrain $C_g(x)$ (including the genuine twist-three contributions) from other spin-independent exclusive processes, see, e.g., \cite{Boussarie:2018zwg}.  Based on the analytical results, we have performed a detailed numerical analysis in both dijet and semi-inclusive diffractive deep inelastic scattering processes. The latter may offer experimental advantages as it does not require the reconstruction of dijet transverse momenta; instead, it involves the inclusive tagging of one or more hadron species.

We have also undertaken the first exploration of quark-channel contributions to double spin asymmetry, revealing an unexpected breakdown of factorization. Indeed, even in the gluon case, we encountered an `end-point' singularity at $x=\pm \xi$ which we circumvented by restricting to $z\sim 1/2$.  
In the quark case, we observed a similar singularity across all $z$ values due to the $C$-odd exchange in the $t$-channel. 
We recognize the importance of revisiting the present  calculation using a dedicated $k_T$ factorization approach. This approach holds promise to remedy the problems with factorization and make the amplitude manifestly finite. But the connection to parton OAMs becomes necessarily less direct.

Overall, our work contributes significantly to the ongoing efforts aimed at unraveling the orbital angular momentum of partons and its role in shaping the spin structure of the nucleon. 
We anticipate that our findings will serve as a valuable foundation for inspiring and guiding further studies in this field.

\begin{acknowledgements}
We thank Saad Nabeebaccus and Feng Yuan for discussion. 
The work of S.~B. has been supported by the Laboratory Directed Research and Development program of Los Alamos National Laboratory under project number 20240738PRD1.
S.~B. has also received support from the U.~S. Department of Energy through the Los Alamos National Laboratory. Los Alamos National Laboratory is operated by Triad National Security, LLC, for the National Nuclear Security Administration of U.~S. Department of Energy (Contract No. 89233218CNA000001).
S.~B. and Y.~H. were supported by the U.~S. Department of Energy under Contract No. DE-SC0012704, and also by  Laboratory Directed Research and Development (LDRD) funds from Brookhaven Science Associates. Y.~H. is also supported by the framework of the Saturated Glue (SURGE)
Topical Theory Collaboration.
\end{acknowledgements}


\appendix

\section{Compton form factors}
\label{a1}
In this appendix we collect formulas for the Compton form factors introduced in (\ref{h2}),  (\ref{l1}) and (\ref{tildemoment}) that are useful for the numerical evaluation of (\ref{main}). 
\begin{align}
{\cal H}^{(1)}_g(\xi) & =\frac{1}{\xi}\left(H_g(\xi,\xi)\ln\frac{1-\xi}{1+\xi}
+ \int_{-1}^1dx\frac{ H_g(x,\xi) - H_{g}(\xi,\xi)}{x-\xi} \right) -i \dfrac{\pi}{\xi} H_g(\xi,\xi), \\[0.2cm]
{\cal H}^{(2)}_g(\xi) & =\dfrac{1}{2} \bigg ( H'_g(\xi,\xi) \ln \dfrac{1-\xi}{1+\xi} + \int_{-1}^1dx\frac{ H'_g(x,\xi) - H'_{g}(\xi,\xi)}{x-\xi} \bigg )\nonumber \\[0.2cm]
& - \dfrac{1}{2\xi} \left(H_g(\xi,\xi)\ln\frac{1-\xi}{1+\xi}
+ \int_{-1}^1dx\frac{ H_g(x,\xi) - H_{g}(\xi,\xi)}{x-\xi} \right) - i\frac{\pi}{2} \bigg (  H'_g (\xi,\xi) - \dfrac{1}{\xi} H_g (\xi,\xi) \bigg ), \label{a2} \\[0.2cm]
\tilde{\cal H}^{(2)}_g(\xi) & =\dfrac{1}{2} \bigg ( \tilde{H}'_g(\xi,\xi) \ln \dfrac{1-\xi}{1+\xi} + \int_{-1}^1dx\frac{ \tilde{H}'_g(x,\xi) - \tilde{H}'_{g}(\xi,\xi)}{x-\xi} \bigg ) -i\frac{\pi}{2\xi}\tilde{H}'_g(\xi,\xi), \\[0.2cm]
{\cal L}^{(2)}_g(\xi) 
& = \frac{1}{2\xi} \bigg (Y_g(\xi,\xi)\ln\frac{1-\xi}{1+\xi} + \int_{-1}^1dx\frac{Y_g(x,\xi)-Y_g(\xi,\xi)}{x-\xi} \bigg ) -i \dfrac{\pi}{2\xi} Y_g(\xi,\xi),
\end{align}
where we defined $H'_g(x,\xi) \equiv \dfrac{d}{dx}{H_g(x,\xi)}$, $\tilde{H}'_g(x,\xi)\equiv\dfrac{d}{dx}{\tilde{H}_g(x,\xi)}$, and 
 $Y_g(x,\xi)\equiv\dfrac{d}{dx} xL_g(x,\xi)$. $C_g^{(2)}(\xi)$ is obtained via the same formula as (\ref{a2}) with $H_g(x,\xi)\to xC_g(x,\xi)$.

Note that, when $\xi\ll 1$, ${\cal H}_g^{(1,2)}$ are dominated by the imaginary parts
\beq
{\mathfrak Im}{\cal H}_g^{(1)}(\xi)= -\frac{\pi}{\xi}H_g(\xi,\xi), \qquad {\mathfrak Im} {\cal H}^{(2)}_g(\xi) = -\frac{\pi}{2} \left(\left.\frac{dH_g(x,\xi)}{dx}\right|_{x=\xi}-\frac{1}{\xi}H_g(\xi,\xi)\right).
\eeq
Moreover, 
assuming a Regge-like behavior $H_g(x,\xi)\sim \frac{1}{x^c}$ with $c\propto \alpha_s$, we see that the second term in ${\cal H}_g^{(2)}$ is parametrically larger than the first term. This leads to the  relation 
\beq
{\rm Im} {\cal H}_g^{(1)}(\xi)\approx -2{\rm Im}{\cal H}_g^{(2)}(\xi). \label{2times}
\eeq

\begin{figure}[htbp]
\includegraphics[width=0.4\linewidth]{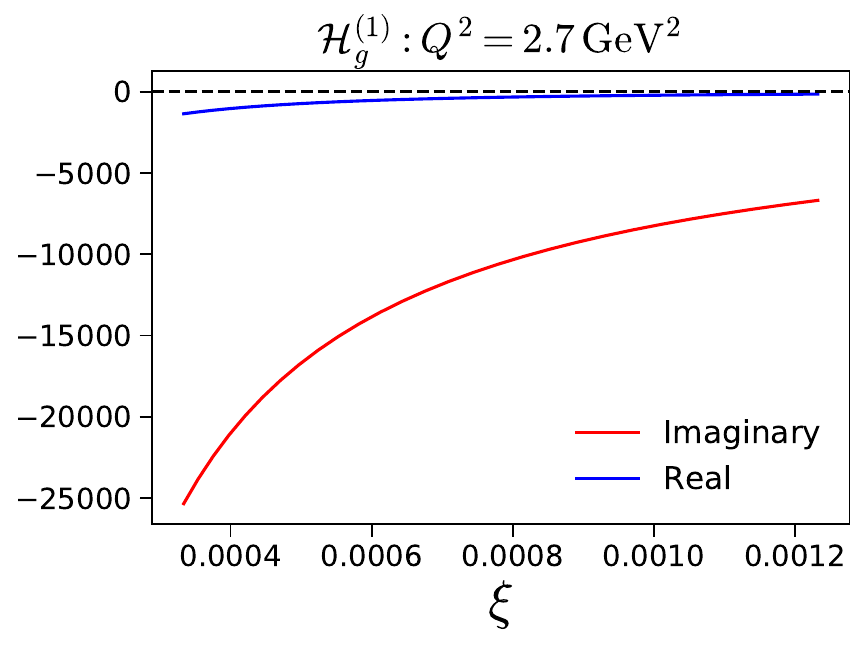}
\includegraphics[width=0.4\linewidth]{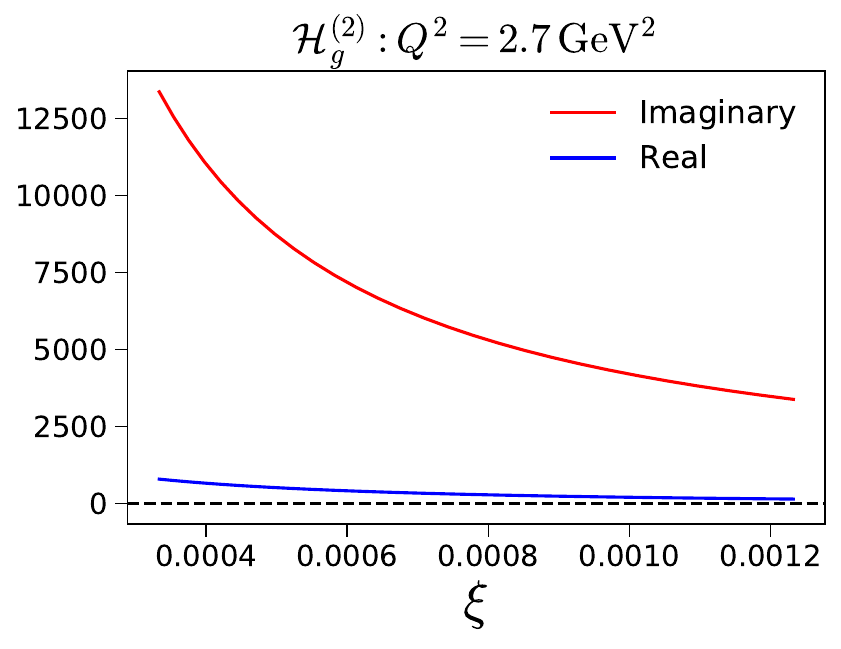}
\caption{CFFs of the unpolarized gluon GPDs.}
\label{cff1}
\end{figure}

\begin{figure}[htbp]
\includegraphics[width=0.4\linewidth]{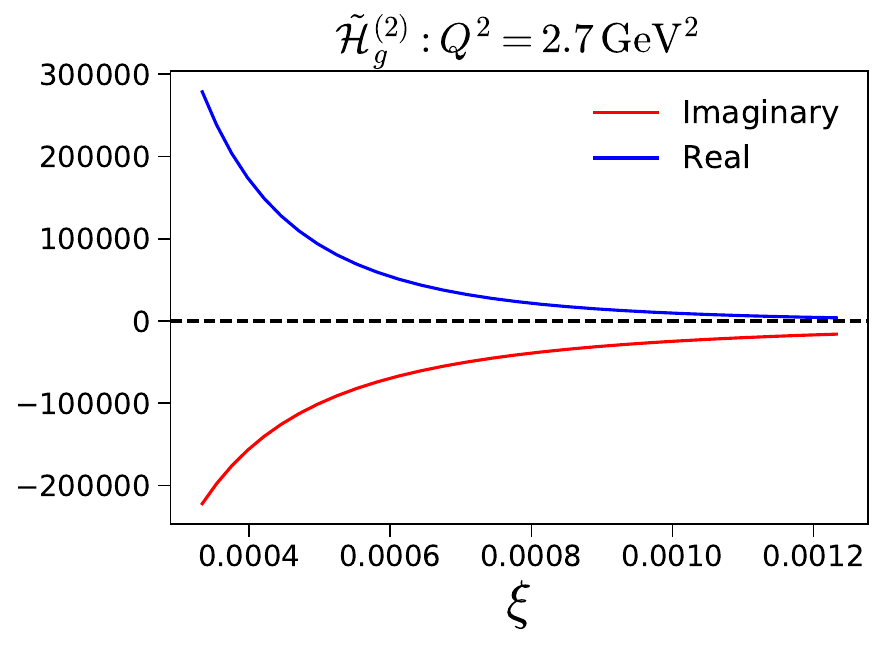}
\includegraphics[width=0.4\linewidth]{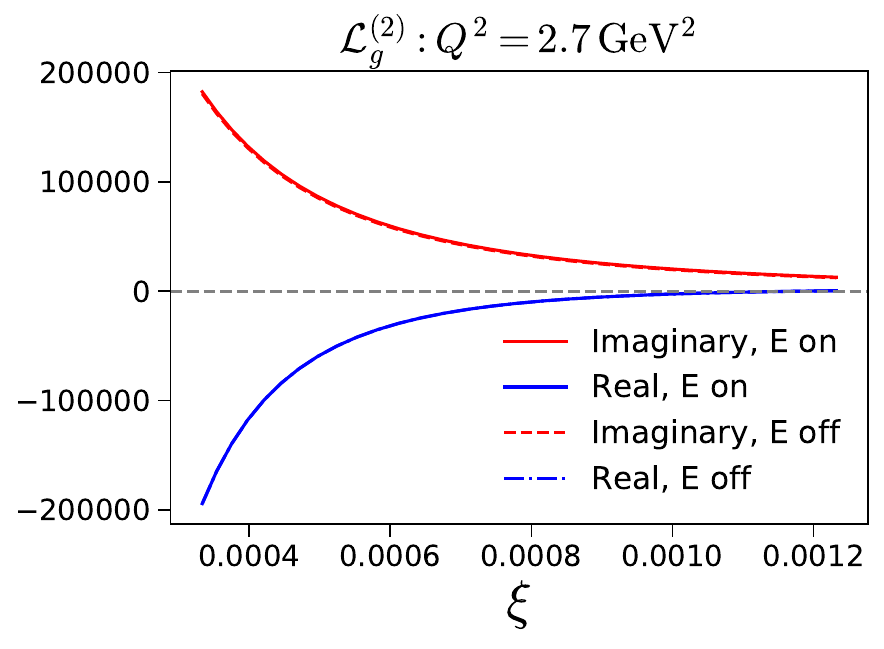}
\caption{CFFs of the helicity (left) and OAM (right) distributions. 
}
\label{cff2}
\end{figure}

\begin{figure}[htbp]
\includegraphics[width=0.4\linewidth]{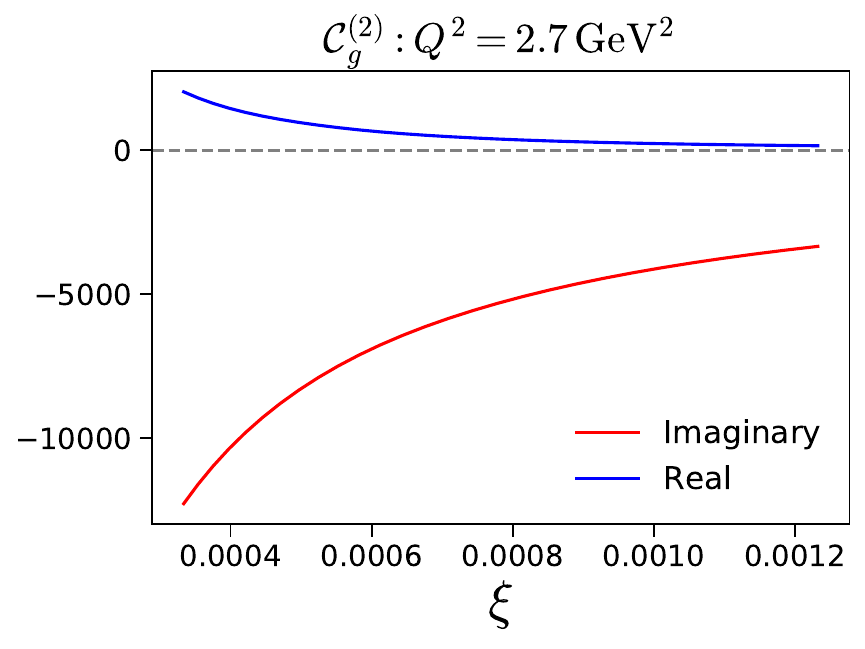}
\caption{CFFs of the spin-orbit correlation.}
\label{cff3}
\end{figure}

In Figs.~\ref{cff1}, \ref{cff2}, and \ref{cff3} we present plots depicting both the real and imaginary components of the Compton form factors as functions of $\xi$. As anticipated, the imaginary components of the Compton factors ${\cal{H}}^{(1,2)}_g$ significantly outweigh their real counterparts at small $\xi$ values. We also see that the relation (\ref{2times}) is approximately satisfied.  However, these properties do not hold for the helicity and OAM Compton factors ${\cal{\tilde{H}}}^{(2)}_g$ and ${\cal L}_g^{(2)}$. From the plots, we observe that the real parts are nearly as prominent as the imaginary ones for small $\xi$ values and they roughly satisfy 
\beq
\tilde{\cal H}_g^{(2)}\approx -{\cal L}_g^{(2)}, 
\eeq
as expected (see a comment after (\ref{dsasimp})). 
Somewhat unexpectedly, however, we find that $|\tilde{\cal H}_g^{(2)}|\sim |{\cal L}_g^{(2)}|$ are about an order of magnitude larger than $|{\cal H}_g^{(1,2)}|$, cf., a comment below (\ref{bfkl}).


\bibliography{ref}
\end{document}